\documentclass[12pt,3p,authoryear]{elsarticle}
\usepackage[T1]{fontenc}
\usepackage{amsmath,amssymb,amsfonts}
\usepackage{natbib}
\usepackage{graphicx}
\usepackage[hyphens]{url}
\usepackage[table]{xcolor}
\usepackage{booktabs}
\usepackage{rotating}
\usepackage{times}
\usepackage{rotfloat}
\usepackage{bm}
\usepackage{xspace}
\usepackage{color}
\usepackage{subdepth}
\usepackage{float}
\usepackage{setspace}
\usepackage{needspace}   
\usepackage[hang,labelfont=bf,position=top]{caption}
\floatstyle{plaintop}
\restylefloat{table}
\restylefloat{figure}			
\journal{The Journal of Financial Markets}

\newcommand{\MEASURE}[3]{$\textit{#1}\mbox{\footnotesize(#2,#3)}$}
\newcommand{\CAT}[2]{\MEASURE{CAT}{#1}{#2}}
\newcommand{\RCAT}[2]{\MEASURE{RCAT}{#1}{#2}}
\newcommand{\CAR}[2]{\MEASURE{CAR}{#1}{#2}}
\newcommand{\CAST}[2]{\MEASURE{CAST}{#1}{#2}}
\newcommand{\NPRCAT}[2]{\MEASURE{NPRCAT}{#1}{#2}}
\newcommand{\NWRCAT}[2]{\MEASURE{NWRCAT}{#1}{#2}}

\newcommand{\insertfloat}[1]{%
\begin{center}
[Insert~#1 about here.]%
\end{center}%
}

\begin{document} 
\singlespacing
\begin{frontmatter}
\title{Media abnormal tone, earnings announcements,\\and the stock market\tnoteref{label1}}
\tnotetext[label1]{We thank the editor (Tarun Chordia), an anonymous referee, Francesco Audrino, Marie-Claude Beaulieu, Samuel Borms, Tolga Cenesizoglu, Serge Darolles, Kilian Dinkelaker, Michel Dubois, George Hübner, Alexandre Jeanneret, Tim Kroencke, Marie Lambert, Gaëlle Lefol, Louis Mangeney, Hannes Mohrschladt, Carlos Ordás Criado, Christophe Perignon, Jérôme Taillard, James Thewissen, Philip Vervliet, seminar participants at  Florence University, HEC Montréal, Skema Business School, and the University of Delaware, as well as participants at AFFI (Nantes, 2021, and Québec, 2019), CFE (Pisa, 2018, and London, 2019), and R/Finance (Chicago, 2018) conferences for helpful comments. We acknowledge the Flemish Science Foundation (\url{https://fwo.be}), Innoviris (\url{https://innoviris.brussels}), IVADO (\url{https://ivado.ca}), and the Swiss National Science Foundation (\url{https://www.snf.ch}, grants \#179281 and \#191730) for their financial support.}
\author[hec]{David Ardia}
\ead{david.ardia@hec.ca}
\author[sher]{Keven Bluteau\corref{cor1}}
\ead{keven.bluteau@usherbrooke.ca}
\author[ghent,vub,vua]{Kris Boudt}
\ead{kris.boudt@ugent.be}
\address[hec]{GERAD \& Department of Decision Sciences, HEC Montréal, Montréal, Canada}
\address[sher]{Department of Finance, Université de Sherbrooke, Canada}
\address[ghent]{Department of Economics, Ghent University, Belgium}
\address[vub]{Solvay Business School, Vrije Universiteit Brussel, Belgium}
\address[vua]{School of Business and Economics, Vrije Universiteit Amsterdam, The Netherlands\\[1cm]
\large Published in Journal of Financial Markets\\
\url{https://doi.org/10.1016/j.finmar.2021.100683}}
\cortext[cor1]{Corresponding author. Université de Sherbrooke, Department of Finance, 2500, boulevard Université, Sherbrooke (Québec) J1K 2R1, Canada. Phone: +1 819 821‑8000.}

\begin{abstract}
We conduct a tone-based event study to examine the aggregate abnormal tone dynamics in
media articles around earnings announcements. We test whether
they convey incremental information that is useful for price discovery for non-financial S\&P 500 firms. The relation we find between the abnormal tone and abnormal returns suggests that media articles provide incremental information relative to the information contained in earnings press releases and earnings calls. 
\end{abstract}
\begin{keyword}
Abnormal returns, abnormal tone, earnings announcements, event study, news media, sentometrics, Structural Topical Model (STM)\\[.2cm]
\end{keyword}
\end{frontmatter}

\clearpage
\setcounter{page}{1}
\doublespacing
\setcounter{page}{2}

\section{Introduction}

An earnings press release provides investors with new information about a firm. This happens directly, through the publication of earnings numbers, as well as indirectly, by triggering qualitative discussions of a firm's prospects, such as through earnings calls and media reports. Current evidence suggests that the media is a double-edged sword for financial markets. On the one hand, the media can reduce information asymmetry by quickly and broadly distributing relevant information, leading to a more efficient market \citep[e.g.,][]{fang2009media,tetlock2010does}. On the other hand, the media can produce short-term price distortions, as news reports can exacerbate the effects of investor biases \citep[e.g.,][]{Tetlock2007,dougal2012journalists,garcia2013sentiment,shiller2017narrative}.

\begin{figure}[H]
\centering
\singlespacing
\caption{\textbf{Information flow}\\
This chart illustrates in a stylized way the information flow between firms, media, and investors during an earnings announcement event.}
\includegraphics[width=1.0\textwidth]{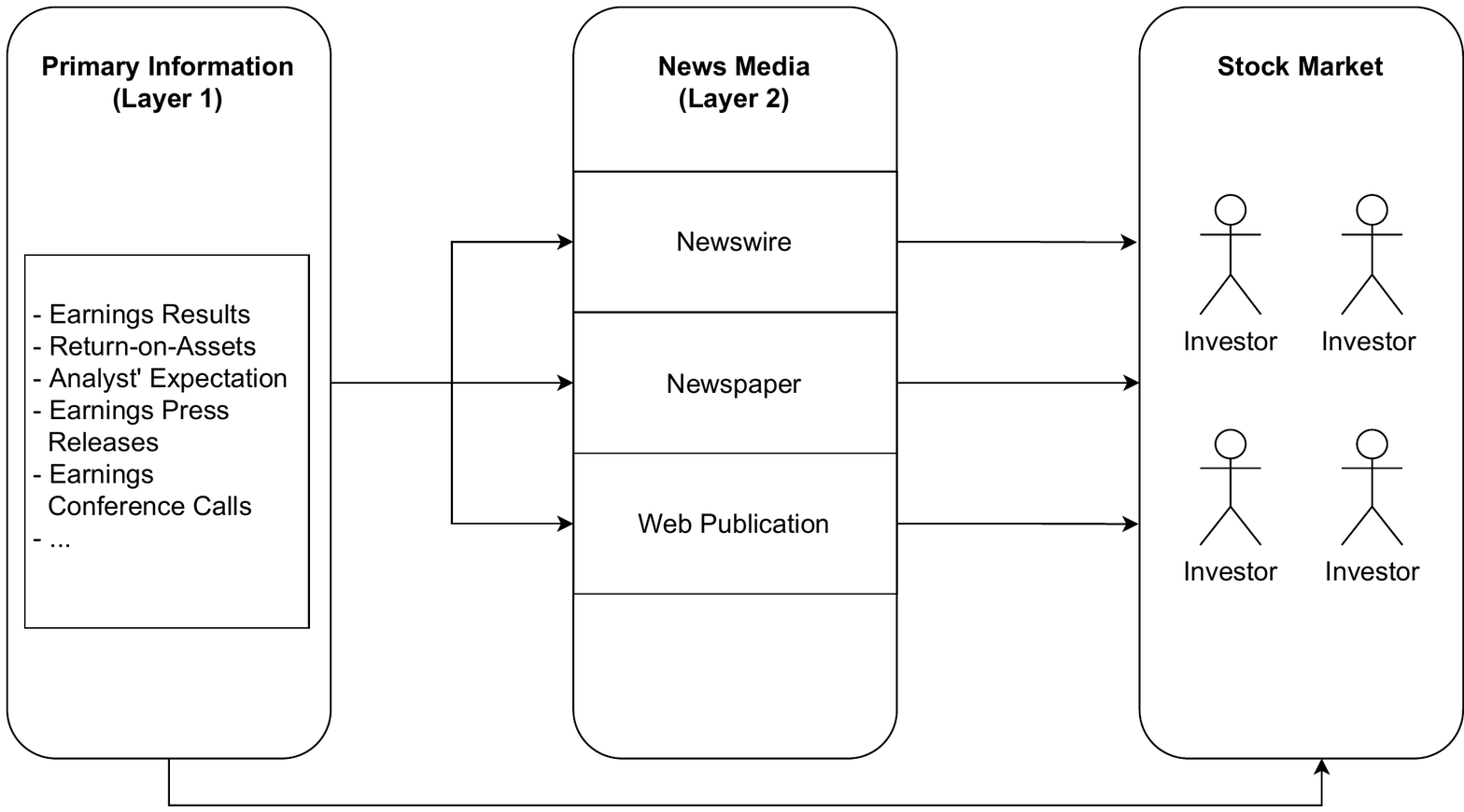}
\label{fig:information}
\end{figure}

We use a novel tone-based event study to evaluate how media reports around earnings announcements and how incremental information provided by the media can help explain stock market reactions around earnings announcements. To put these research questions into context, Figure~\ref{fig:information} displays a flowchart illustrating in a stylized way the information flow between firms, the media, and investors. According to the Regulation Fair Disclosures (Reg FD) in the United States, publicly traded firms are obliged to disclose all material non-public information. In the case of earnings announcements, this is done via earnings calls and earnings press releases, as illustrated in the ``primary information'' layer of Figure~\ref{fig:information}. Through those channels, firms provide media and investors with quantitative and qualitative information about their financial results and business prospects.

When the information is of interest to their readers, news media outlets then report about the earnings event, often contextualizing the information and adding analysts' comments and forecasts. As formalized by \citet{NimarkPitschner2019}, the media thus act as information intermediaries between investors and the true state of the world of the firm \citep[see also][]{LarsenEtAl2021}. The second layer in Figure~\ref{fig:information} relates to this indirect transmission, where the media disseminate and transform the first layer of information to investors. Under this framework, we refer to dissemination when the media transmit the most important information in the first layer, whereas transformation corresponds to any modification, completion, or contextualization of the information available in that layer. 

\insertfloat{Figure~\ref{fig:information}}

As a concrete example of the dissemination and transformation of the media, consider the earnings announcement of Home Depot 
on August 19, 2008. The firm reported earnings per share of 71 cents that beat the average analyst forecast of 61 cents. One of the news articles 
in our corpus reporting about that event is published by CNN with the title ``Home Depot profit tumbles 24\%.'' In that article, the journalist disseminates the 
information from the earnings press release and earnings call about the higher-than-expected earnings and its relation to the stimulus checks sent by the government in May. Moreover, the journalist transforms the first layer of information by referring to another article published by CNN stating that ``(...) sales dropped off in July, suggesting the program had run its course.'' Hence, the increase explained by the stimulus checks is not deemed sustainable.\footnote{The two news article published by CNN are available at \url{https://money.cnn.com/2008/08/19/news/companies/home_depot_earns/index.htm?postversion=2008081906} and \url{https://money.cnn.com/2008/08/13/news/economy/retail_sales/index.htm?postversion=2008081311}, the earnings press release at \url{https://ir.homedepot.com/news-releases/2008/08-19-2008}, and the earnings call transcript at \url{https://seekingalpha.com/article/91649-home-depot-f2q08-qtr-end-8-3-08-earnings-call-transcript?part=single}.}

The dissemination and transformation of the media are reflected in the tone. Based on the above framework, we posit two hypotheses. First, the media tone provides a complementary signal about the firm compared to what can be predicted using only the 
information derived from the earnings report, the earnings press release, and the earnings call. The complementary role of the news media stems from their transformation of earnings-related information into a more easily understandable, contextualized, and condensed format, as well as the additional information it can provide given important contemporary events. Second, controlling for the information contained in earnings press releases and earnings calls, we expect that the tone of media articles is useful for stock price discovery following an earnings announcement. This hypothesis implies that the media contributes to disseminating and transforming information to investors during earnings announcements.

To test this empirically, we conduct a novel tone-based event study. We generalize the data-driven word power methodology of \citet{jegadeesh2013word} to the setting of multiple news articles per day to compute the daily media tone for firms.\footnote{\citet{jegadeesh2013word} quantify the tone of words in 10-K fillings  by means of a regression of the firm's stock return on the presence of sentiment words in the corresponding 10-K. They thus have one document per observation period while we have multiple.} Then, we proceed in two steps. First, for a given firm and time, we define an ``abnormal tone'' as an unexpected tone conditional on a market capitalization-weighted average of the media tone about firms in the universe. The abnormal tone is designed to capture the idiosyncratic part of the news as opposed to news generated by general market conditions. We aggregate the abnormal tone into a ``cumulative abnormal tone'' (CAT) measure around the event. Then, we filter the CAT to remove the information that can be derived from earnings press releases and earnings calls, as well as other variables such as book-to-market ratio and firm size. We refer to this incremental information measure as the ``residual cumulative abnormal tone'' (RCAT).\footnote{\citet{huang2013tone} pioneer the approach of analyzing incremental tone by means of the residuals of a regression. They do this for earnings press releases and interpret the residual tone as the tone management by the firm. It is inspired by the calculation of discretionary accruals in accounting and finance.}  We apply our approach to media articles on non-financial S\&P 500 firms' quarterly earnings announcements from 2000 to 2016. In doing so, we analyze more than 2.3 million news articles published by newswires, newspapers, and web publications. 

We first analyze the drivers of CAT around earnings announcements. As expected, we find that CAT is largely driven by the information disclosed by the firm, such as the sign of the earnings surprise, the tone of the press release, and the tone of the earnings call. CAT is also driven by its past values, suggesting that the new information embedded in CAT also incorporates past information.  However, a large part of the CAT's variance remains unexplained, which supports the use of RCAT to analyze incremental information from media tone. 

We then investigate whether RCAT conveys incremental information about a firm's stock price reaction around the announcement. We find a significant positive contemporaneous relation: an increase to RCAT of one standard deviation is associated with an expected excess abnormal return of about 0.84\%. Furthermore, we find that RCAT has predictive power for future price movements. Specifically, we find a partial reversal in stock price movements associated with RCAT in the month following the announcement; a one standard deviation increase in RCAT is associated with an excess abnormal return of -0.36\%. One possible channel is that stock markets overreact to incremental information conveyed by the media's abnormal tone. This interpretation is consistent with our finding that the absolute value of RCAT is positively related to a firm's abnormal trading volume around the earnings announcement.

Our paper contributes to two different strands of research. First, it advances the literature on the quantification of information available in textual data to better understand the role of the news media in financial markets. Early pioneers in this literature include \citet{engelberg2008costly} and \citet{tetlock2008more}, who used the General Inquirer \citep{stone1963computer} system to compute the sentiment of media publications such as \emph{Wall Street Journal} columns and newswire articles. Since \citet{LoughranMcdonald2011}, it has become best practice to use a domain-specific lexicon in accounting and finance. While several lexicons exist to link the stock price reaction to the textual information in corporate disclosures, there is no such lexicon for media analysis. As also advocated by \citet{ke2019predicting}, the use of a data-driven calibration of a lexicon fills that gap. In this study, we propose an extension of the word power methodology proposed by \citet{jegadeesh2013word} to obtain a domain-specific lexicon for analyzing the daily tone of firms in media articles with the goal of explaining the stock market. We then develop a comprehensive
event-study framework designed to isolate the information from the daily tone that is incremental to other sources of information. We use a structural topic model (STM) approach \citep{RobertsEtAl2016} to analyze the difference in content between the three types of sources: media news articles, earnings press releases, and earnings calls. Compared to the traditional topic model, the STM has the advantage that the effects of metadata can be modeled. We exploit this feature to provide insights on how the media differentiate themselves from earnings press releases and earnings calls by focusing on certain subjects. As such, we go beyond the analysis of tone in studying the mechanism through which the media provide additional information to the market. Such a topic analysis provides insights on how the media focus on certain subjects of discussion to contextualize, emphasize, or mitigate specific results compared to earnings press releases and earnings calls. While the focus of our study is on earnings announcements, our analysis can easily be transposed to analyze the role of media in transforming the information around other events, such as central bank announcements, natural disasters, or other corporate events.

Second, the results contribute to the vast literature on price discovery around corporate disclosure events, and in particular, earnings announcements. Several studies focus on the information contained in the tone of newswire articles \citep{engelberg2008costly}, the tone of earnings press releases \citep{demers2008soft,Henry2008,davis2012beyond,huang2013tone,arslan2016managers}, the tone of earnings calls \citep{PriceEtAl2012,BorochinEtAl2018}, the narrative structure of voluntary disclosure narratives \citep{allee2015structure},  as well as their level of readability \citep{jones1994accounting,li2008annual,merkley2014narrative,asay2018firm}. Whether markets overreact or underreact to the abnormal tone in media reports is still an open question. \citet{engelberg2008costly} analyzes only the Dow Jones Newswire publications and finds an underreaction that is only economically significant for the 40 to 80 days following the announcement.\footnote{See Figure 2 in \citet{engelberg2008costly}.} When the tone of media articles captures the attention of retail investors, an overreaction is expected by most articles in the literature on investors' attention bias \citep[e.g.,][]{da2011search}. We test whether markets overreact or underreact to media abnormal tone using a comprehensive corpus of news from several newswire, newspaper, and web publication sources. Our findings confirm that there is an overreaction to the abnormal tone of media news articles leading to a price reversal in the 20 days following the event. 

This paper is organized as follows. Section~\ref{sec:metho} presents our tone-based event study. Section~\ref{sec:data} describes our data. Section~\ref{sec:catdriver} presents the empirical results on the media coverage and abnormal tone around earnings announcements. Section~\ref{sec:catcar} examines the media tone and the stock market reaction around earnings announcements. Finally, Section~\ref{sec:conclusion} concludes.

\section{Tone-based event study}
\label{sec:metho}

\citet{algaba2016econometrics} define tone as ``the disposition of an entity toward an entity expressed via a certain medium.''\footnote{Other popular terms are ``sentiment,'' ``polarity,'' or ``opinion.'' Under the tone implementation we employ, the tone can be interpreted as the ``expected return implied by textual documents.'' Throughout the paper, we use ``tone'' for two reasons. First, it is the naming convention used in \citet{jegadeesh2013word}, which is a reference in the field. The authors use the word ``tone'' more than a hundred times in the context of 10-Ks, which are also factual documents. Second, our event-study framework can be used with alternative methods (supervised or not) than the word power to extract the tone, which do not necessarily lead to an interpretation as ``expected return implied by textual documents.''} This disposition can originate from facts (i.e., the firm's EPS beats analyst consensus), referred to as factual sentiment by \citet{HeerschopEtAl2011}, or opinions (i.e., the firm announced a disappointing quarter, but discussed an innovative and impressive new product line). \citet{Kim2019}, for instance, differentiate between factual tone and promotional tone. Given the difficulty of identifying facts from opinions in an automated
way, we do not differentiate between them in this paper. 

Below, we present our approach to extract the (abnormal) tone about firms from a sample of media articles published close to earnings announcements (regardless of whether the articles discuss earnings results). Our implementation follows \citet{jegadeesh2013word} and leads to tone estimates that can be interpreted as the expected stock return implied by the textual documents. While this follows strictly the naming convention used by \citet{jegadeesh2013word}, our event-study approach is agnostic of the tone computation method. It can be implemented with various alternative methods (supervised or not) than the approach of \citet{jegadeesh2013word}. 

\subsection{Measurement of the tone}

We follow \citet{jegadeesh2013word} in using a data-augmented bag-of-words approach to measure the tone in news articles about firms. The calibration methodology of the lexicon, detailed in the Internet Appendix, Section~A, leads to a dictionary of $J$ words with associated polarity scores $\zeta_{j}$, $j=1,\ldots,J$, where the scores set the polarity of each word (i.e., the degree of positivity or negativity). For a firm $k = 1, \dots, K$, the tone at time $t$  is defined as the weighted average of the polarity scores of the words in the text:
\begin{equation}
\textit{TONE}_{k,t} = \sum_{j=1}^{J}\zeta_{j}f(j,k,t) \,,
\end{equation}
where $f(j,k,t)$ maps a word $j$  for firm $k$ at time $t$ to a real number. 

We define the function $f$ as an average of term frequencies in news articles published at time $t$. More precisely, let $D_{k,t}$ be the number of articles written about firm $k$ at time $t$, conditional on the presence of at least one  word $j$ in each article $d = 1,\dots,D_{k,t}$, $\textit{FQ}_{d,j,k,t}$ as the number of times the $j$th word is encountered in article $d$, and $N_{d, k,t}$ as the number of words in article $d$. Then, the average term-frequency for word $j$, firm $k$, at  time $t$ is:
\begin{equation}
f(j,k,t)  = \frac{1}{D_{k,t}}\sum^{D_{k,t}}_{d = 1}\frac{\textit{FQ}_{d,j,k,t}}{N_{d,k,t}}\,.
\end{equation}
The word power definition of $f(j,k,t)$ from \citet{jegadeesh2013word} is a special case of our definition where $D_{k,t} = 1$ at all times for all firms. Due to the flexibility in defining  $f(j,k,t)$, we refer to our methodology  as the generalized word power (GWP).  

\subsection{Definition of the cumulative abnormal tone}

Given an event-firm  $i$ and firm-related articles published at time $\tau$ relative to the earnings announcement date $t_i$, we decompose tone into the sum of a ``normal tone'' and an ``abnormal tone'':
\begin{equation}\label{eq:tonedecomposition}
\textit{TONE}_{i,\tau} = \textit{NTONE}_{i,\tau} + \textit{ATONE}_{i,\tau}\,,
\end{equation}
where the relative time $\tau$ is zero for the earnings announcement day. We use this decomposition of the observed tone to extract the 
new information as revealed in the abnormal tone.  This interpretation is motivated by the fact that, in the hypothetical case where the media predominantly publish information already known or expected, the abnormal tone should be much closer to zero than if the media just published new information.\footnote{\citet{huang2013tone} and \citet{arslan2016managers} use a measure of abnormal tone for firms' earnings press releases where the normal tone is modeled using firms' fundamentals. They define the abnormal tone as a measure intended to capture the discretionary and inflated component of tone. We later use this method to extract the media's transformation of information about earnings announcements.}

Finally, to evaluate the effect of an earnings announcement on the abnormal tone over a time window of $\tau_1$ to $\tau_2$ ($\tau_1 < \tau_2$) relative to the event day, we define the cumulative abnormal tone, CAT, as follows:
\begin{equation}
\textit{CAT}_i\mbox{\footnotesize{($\tau_1$,$\tau_2$)}}
= 
\sum_{\tau = \tau_1}^{\tau_2} \textit{ATONE}_{i,\tau}\,.
\end{equation}

\subsection{Estimation of the normal tone model}

The decomposition of tone into normal and abnormal components in \eqref{eq:tonedecomposition} requires an estimate of the normal tone. For this purpose, we use a linear factor model:
\begin{equation}\label{eq:excesssent}
\textit{TONE}_{i,\tau} = \alpha_i + \beta_{i} F_\tau + \epsilon_{i,\tau}\,,
\end{equation}
where $F_{\tau}$ is a tone factor common across all firms at relative time $\tau$; $\alpha_i$ and $\beta_i$ are the event-firm constant and factor exposure, respectively; and $\epsilon_{i,\tau}$ is an error term. This specification is analogous to the model of expected returns commonly used in abnormal-return event studies \citep[e.g.,][]{mackinlay1997event}.\footnote{This model nests the special case of setting the normal tone to an event-specific constant ($\beta_{i} = 0$).} For the tone factor $F_t$, we use a weighted average of the firms' tones for the $K_t$ firms for which we can compute a tone value at time $t$:
\begin{equation}\label{eq:factor}
F_t = \sum_{k = 1}^{K_t} \omega_{k,t} \textit{TONE}_{k,t}\,, 
\end{equation}
where $\omega_{k,t}$ is the firm-$k$ weight  at time $t$ and  $\sum_{k = 1}^{K_t}\omega_{k,t} = 1$. $F_{\tau}$ is then defined as the appropriately aligned value of $F_t$ at relative time $\tau$.  

We estimate the constant $\alpha_{i}$ and the factor exposure $\beta_{i}$ in \eqref{eq:excesssent} by ordinary least squares (OLS) using data available in an estimation window $[t_i - L - K, \dots, t_i - K - 1]$, where $L$ is the length of the window and $K$ is the offset relative to the event date $t_i$. The abnormal tone is then measured over an event window:
\begin{equation}\label{eq:nt}
\textit{ATONE}_{i,\tau} = \textit{TONE}_{i,\tau} - (\widehat{\alpha}_{i} + \widehat{\beta}_{i} F_{\tau} ) \,,
\end{equation}
for $\tau > t_i - K - 1$, where $\widehat{\alpha}_{i}$ and $\widehat{\beta}_{i}$ are the estimated event-entity-specific constant and factor exposure, respectively. 

\begin{figure}[H]
\centering
\singlespacing
\caption{\textbf{Abnormal tone event study timing information}\\
This figure shows the GWP tone estimation scheme (Panel~A) and the event study methodology (Panel B). The GWP model is trained at the end of each year using an expanding window (with the sample data range indicated on the horizontal lines). Then, each GWP-trained model is used to compute the tone of news articles for the following year (OOS text boxes). Panel B is based on the Home Depot earnings announcement event for the second quarter of 2008. We use an estimation window of $L = 30$ days, with an offset of $K = 5$ days, and an event window of 20 days. ``TD'' stands for trading days.}
\includegraphics[width=1\textwidth]{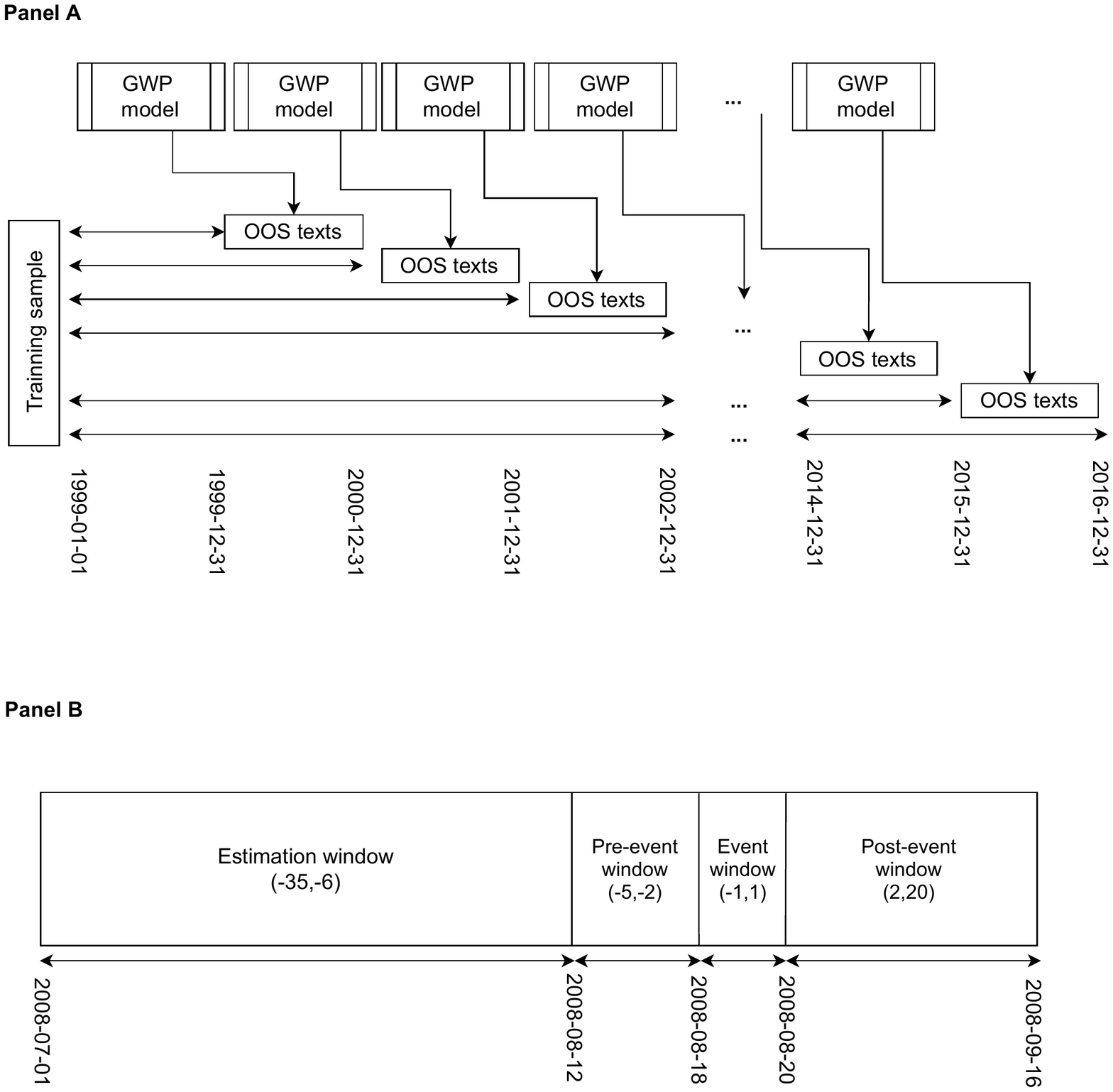}
\label{fig:timing}
\end{figure}

For illustration, Figure~\ref{fig:timing} shows how our methodology applies to Home Depot's 2008 second-quarter earnings announcement (August 19, 2008). First, we collect all relevant news articles published about Home Depot for days surrounding August 19, 2008. Second, for each day, we measure the daily tone extracted from the news articles using the GWP model calibrated on data observed up to December 31, 2007. Third, we estimate the normal tone model with $L = 30$ and $K = 5$ days. Finally, we compute the abnormal tone and CAT over various post-estimation windows of interest (pre-event, event, and post-event windows).

\insertfloat{Figure~\ref{fig:timing}}

\subsection{Definition of residual CAT}

Our definition of abnormal tone captures the idiosyncratic part of the news as opposed to news generated by general market conditions. We can go one step further and isolate or filter out a specific part of the new information. In particular, in the case of scheduled news such as earnings announcements, we can remove the effect of the firm's communication and financial results from the media's reporting to assess how the media transforms that information. To do so, we follow the procedure outlined in \citet{huang2013tone}. Specifically, we define the residual CAT of  event-firm $i$, $\textit{RCAT}_i\mbox{\footnotesize{($\tau_1$,$\tau_2$)}}$, by the transformation  of  $\textit{CAT}_i\mbox{\footnotesize{($\tau_1$,$\tau_2$)}}$, such that it is orthogonal to a vector $\mathbf{x}_i$ of information originating from first-layer sources (e.g., the firm and market participants):
\begin{equation}
\textit{RCAT}_i\mbox{\footnotesize{($\tau_1$,$\tau_2$)}}
= \textit{CAT}_i\mbox{\footnotesize{($\tau_1$,$\tau_2$)}}
- \widehat{\boldsymbol{\theta}}_{\tau_1,\tau_2}' \mathbf{x}_{i} \,.
\end{equation}	 
The vector of coefficients $\widehat{\boldsymbol{\theta}}_{\tau_1,\tau_2}$ is estimated from the regression $\textit{CAT}_i\mbox{\footnotesize{($\tau_1$,$\tau_2$)}} = \boldsymbol{\theta}_{\tau_1,\tau_2}' \mathbf{x}_{i} + \eta_i$, where $\eta_i$ is an error term. Primary information for earnings announcements include the tone of the earnings press release, the tone of the earnings call, the earnings surprise, and the return on assets. 

\subsection{Residual and abnormal tone of individual news articles}

To investigate a specific news article (or a group of articles), we isolate the contribution of each article to the daily (abnormal) tone. The tone of a single article $d$ discussing event-firm $i$ and published at relative time $\tau$ is obtained as:
\begin{equation}
\textit{TONE}_{d,i,\tau} = \sum_{j=1}^{J}\widehat{\zeta}_{j}\frac{\textit{FQ}_{d,j,i,\tau}}{N_{d,i,\tau}}\,,
\end{equation}
from which we can compute the abnormal tone of a specific news article as:
\begin{equation}
\textit{ATONE}_{d,i,\tau} = \textit{TONE}_{d,i,\tau} - ( \widehat{\alpha}_i + \widehat{\beta}_i F_\tau ) \,.
\end{equation}

Finally, we can measure the article-specific residual abnormal tone  contribution,\linebreak $\textit{RATC}_{d,i,\tau}\mbox{\footnotesize{($\tau_1$,$\tau_2$)}}$, as follows:
\begin{equation} \label{eq:ratc}
\textit{RATC}_{d,i,\tau}\mbox{\footnotesize{($\tau_1$,$\tau_2$)}}
= \frac{\textit{ATONE}_{d,i,\tau}}{D_{i,\tau}} - \frac{\widehat{\boldsymbol{\theta}}_{\tau_1,\tau_2}' \mathbf{x}_{i}}{\sum_{j = \tau_1}^{\tau_2} D_{i,j}}\,,
\end{equation}
where $\tau_1 \leq  \tau \leq \tau_2$.

\section{Data}
\label{sec:data}

Our empirical analysis focuses on news articles discussing 598 non-financial firms that were included in the S\&P 500 Index between 2000:Q1 and 2016:Q4.\footnote{We only consider firms at the time they were included in the S\&P 500 Index. Our sample thus tracks S\&P~500 constituents over time. We exclude the financial sector, as is commonly done in earnings-announcement event studies. Financial firms are identified using the first two digits of the Global Industry Classification Standard code for each firm.}

\subsection{Earnings, accounting, and stock market data}

We collect quarterly earnings dates, values, and analyst forecasts from the I/B/E/S database.\footnote{We obtain forecasts of quarterly figures reported in the quarter prior to a quarterly report's publication (i.e., \textit{FPI} = 6) from the Forecast Period Indicator.}\textsuperscript{,}\footnote{Earnings announcements can be made outside of trading hours. As such, there can be differences in the timing of market reaction to earnings announcements. To account for this, we follow \citet{engelberg2018anomalies}, and redefine each earnings announcement date as the day with the highest trading volume in a three-day window centered on the I/B/E/S earnings announcement date.} We gather the quarterly asset value, net income, book value, market capitalization, daily stock prices, and trading volumes from the merged CRSP-Compustat database. We match the two data sources using the I/B/E/S CUSIP and the Compustat NCUSIP.

To identify ``good'' and ``bad'' earnings announcements, we follow \citet{livnat2006comparing} and compute the analysts' standardized unexpected earnings (also called the ``earnings surprise''):
\begin{equation}
\textit{SUE}_i = \frac{\textit{EPS}_i - \textit{MFOR}_i}{P_i}\,,
\end{equation}
where, for event-firm $i$, $\textit{EPS}_i$ is the reported earnings per share, $\textit{MFOR}_i$ is the median analyst forecast of earnings per share, and $P_i$ is the price of the firm at the end of the earnings quarter. We consider only the most recent forecast for each analyst produced no more than 90 days before an earnings announcement.

To measure the share turnover, we follow \citet{campbell1996measuring} and compute the 
log-percentage of outstanding shares traded: 
\begin{equation}
\textit{ST}_{i,\tau} = \log\left(\frac{V_{i,\tau}}{C_{i,\tau}}\right)\,,
\end{equation}
where $V_{i,\tau}$ is the trading volume and $C_{i,\tau}$ is the number of common shares outstanding, for event-firm $i$ at relative time $\tau$.

\subsection{Textual data}

First, we collect firms' historical company names and tickers from Compustat. For each historical company name, we retrieve all news articles available on LexisNexis with a minimum relevance score of 85.\footnote{LexisNexis indexes each article with metadata information such as the company or topic of the article. These metadata tags are associated with a relevance score indicating whether there is a major reference to the tag. To avoid sampling errors, we manually verify that each historical company name is matched to the LexisNexis equivalent company metadata tag and remove company names without a LexisNexis equivalent company tag.} This database is used, for instance, in \citet{fang2009media} and \citet{ahmad2016media}. Our news sample spans from January 1, 1999, to December 31, 2016.\footnote{We use 1999 data to calibrate the word scores for 2000.} 

We then use the following filters:

\begin{itemize}
\item We only keep news articles categorized by LexisNexis English as a ``newswire,'' ``newspaper,'' or ``web publication.''

\item We eliminate articles with fewer than 200 words, as tone measures for short articles are noisy \citep[see][]{shapiro2017measuring}.

\item We exclude patent announcements by leveraging LexisNexis' topic classification, as they are irrelevant to our study. 

\item We remove duplicate (or near-duplicate) articles using locality-sensitive hashing \citep[see][]{wang2014hashing}.

\item We remove machine-generated texts, such as automatic daily stock picks from newswires. These articles are mainly composed of numbers, are highly structured, and lack opinions because they are not written by individuals. 

\item We remove articles for which there are more than two  firms (or tickers) with a major reference (i.e., relevance score of at least 85) to avoid mixing the tone of several firms \citep*[see][]{kelley2013wise}.

\item We remove the sources Midnight Trader - Live Briefs and News Bites - US Markets, which have a large number of news items (they are in the top 100 sources based on the number of news items), but predominantly provide market summaries and trading recommendations. In most cases, the type of news articles provided by these sources are automatically generated.\footnote{We verified that the conclusions of this study remain qualitatively similar when we remove the last two filters.}
\end{itemize} 

Each news article is processed to remove web address (URL), numbers, punctuation, spaces, and other non-textual material. We then transform all words to their root form using the Porter stemming algorithm  \citep{Porter1980}. More details on the various algorithms and software used in the cleaning process are provided in the Internet Appendix, Section~B.

The resulting database contains earnings press releases and transcripts of the earnings calls (distributed through newswires), and news articles relevant to test our hypotheses on media transformation. It is crucial to split news incoming from the media and news incoming from the firm to evaluate the relation between the media and the market. We rely on rule-based procedures to identify the earnings press releases and the earnings call transcripts in the news database. For press releases, we extract articles from newswires that have been tagged as either ``press releases,'' ``company earnings,'' ``financial results,'' ``interim financial results,'' ``financial performance and reports,'' or ``earnings per share.'' We also require that the articles contain the word ``contact,'' as contact information is typically included in earnings press releases. We then apply a negative filter over the selected newswire articles using keywords associated with misclassifying newswire articles as earnings press releases. Finally, in case of multiple detections of earnings press releases in the newswire data, we select the most likely one based on keyword similarity with a training set of earnings press releases. For earnings calls transcripts, we extract documents classified as ``transcript''  in LexisNexis for each firm. Again, we use keyword similarity selection in case of multiple detections.

\begin{figure}[H]
\centering
\singlespacing
\caption{\textbf{Number of articles over time}\\
This figure shows the number of  news  articles per year by source type from 1999 to 2016.}
\includegraphics[width=1\textwidth]{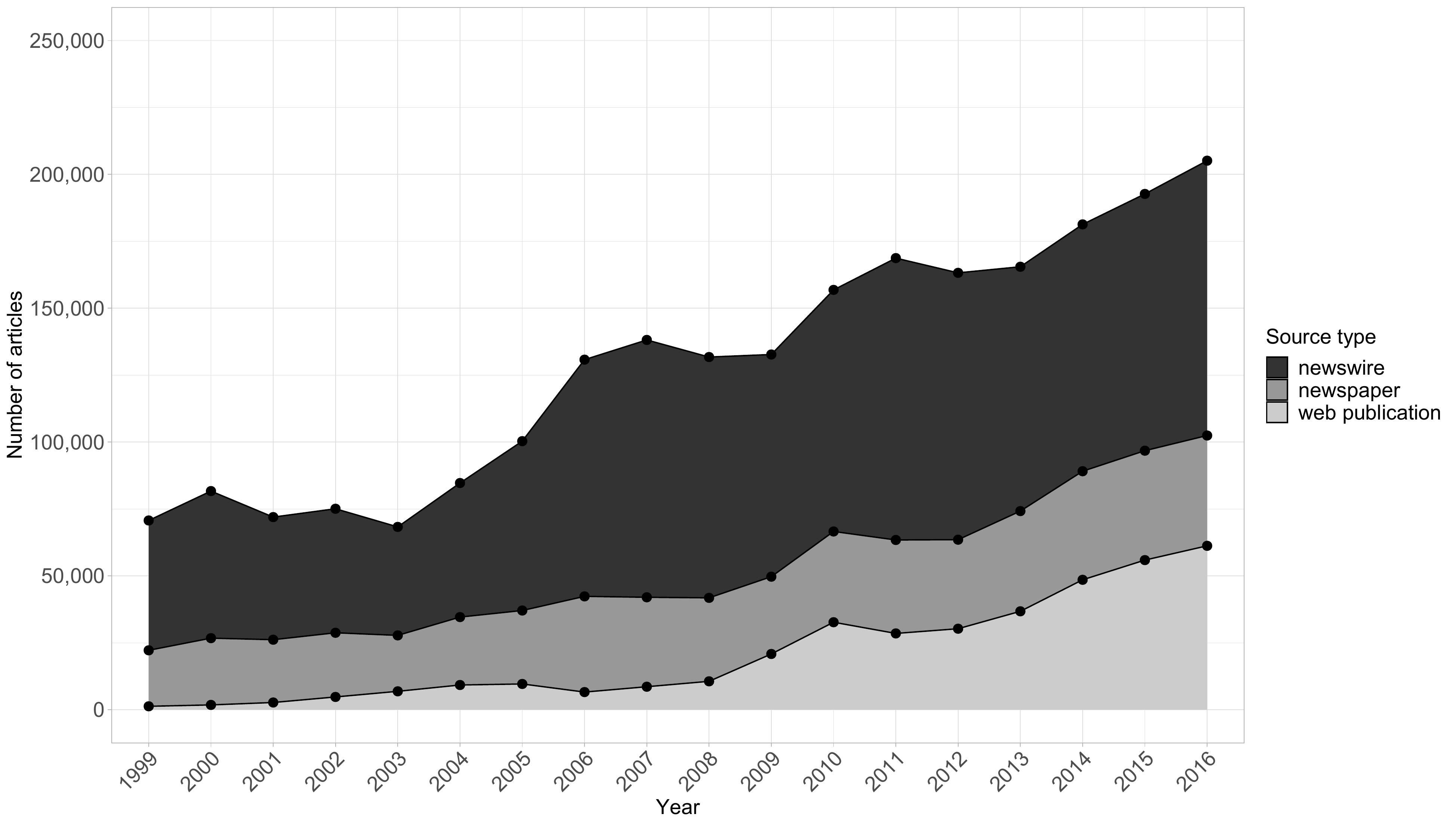}
\label{fig:newsperyear}
\end{figure}

In Figure~\ref{fig:newsperyear}, we display the number of news articles retrieved by publication type over time. Newswires dominate other sources for the entire sample. We also see substantial growth in the number of articles over time. This is due to an increase in the publication frequency and the number of sources available. For example, the number of articles published about S\&P 500 firms originating from web publications is negligible before 2008 but dominates newspapers after 2010.

\insertfloat{Figure~\ref{fig:newsperyear}}

In Table~\ref{tab:coverage}, we report summary statistics on the number of articles and daily coverage. We define the daily coverage as the percentage of trading days with at least one news item discussing the firm. Overall, our sample contains 2,319,582 news articles, with an average of 3,878 articles per firm and an average daily coverage of 34.85\% (i.e., for a typical firm, on about one in every three trading days, there will be at least one article published about the firm; on the other two trading days, there will be no articles). The minimum daily coverage for a firm in our sample is 0.27\%, and the maximum daily coverage is 100\%. We further split the sample across four buckets of 149 firms conditional on the average market capitalization computed for each firm over the 1999 to 2016 period. Firms with the lowest market capitalization have an average of 948 news articles discussing them and an average daily coverage of 20.67\%. Firms with the highest market capitalization have an average of 10,449 news articles and an average daily coverage of 54.82\%. Hence, there is a positive relation between the number of articles (and the daily coverage) and the market capitalization of the firm (at least in the S\&P 500 universe). This relation holds across different types of publications.\footnote{\citet{engelberg2008costly} finds a similar relation for the Dow Jones newswire and posits several reasons for this, such as more expansive analyst coverage, catering to client demand (i.e., a larger shareholder base), or even market capitalization internal thresholds for news coverage.}

\insertfloat{Table~\ref{tab:coverage}}

In the Internet Appendix, Section~C, we list the top sources of news articles, for which we report the source name, the source type, the number of news articles included in our sample, and a brief description of the source and target content. 

\subsection{Tone measurement for news articles}

We now describe how we calibrate our GWP methodology to our data set. To capture the market's reaction to information published by the media, we follow \citet{jegadeesh2013word} and calibrate the word scores on firms' stock returns. This way, we can also interpret the computed textual tone as the expected return implied by the media articles. The list of words in our dictionary is obtained by merging the \citet{LoughranMcdonald2011} and Harvard IV-4 \citep{stone1963computer} lexicons. We use the Porter stemming algorithm \citep{Porter1980} to eliminate noninformative word variations. Our lexicon contains 3,585 root words. 

We train the GWP and measure the tone using an expanding window (see Panel~A of Figure~\ref{fig:timing}). To ensure that the market reaction is concurrent with news articles, we only use newswires to calibrate the GWP, as newspapers and web publications can be delayed \citep{fang2009media}. In addition, to obtain reliable score estimates, we remove words that are observed in fewer than 200 days across all firms for each expanding window. Finally, to avoid any look-ahead bias, we use the calibrated scores to measure the tone of the news articles of the year following the last year of each training window. This filter and estimation scheme results in 722 valid root words in the first estimation window (from 1999 to 2000), and 2,476 valid root words for the last estimation window (from 1999 to 2015). 

Finally, we aggregate the tone of each article into a text-based factor $F_t$, whose weight in \eqref{eq:factor} is proportional to the market capitalization of common shares outstanding for firms for which a tone observation is available at time $t$. Our motivation for this choice is twofold. First, as observed in our sample, large firms (based on market capitalization) tend to have more news articles written about them than small firms. As such, large firms should have a bigger impact on the overall market tone. Second, a market capitalization-weighted text-based normal tone factor mimics the traditional market model used in the event study literature for measuring abnormal returns.\footnote{The results remain qualitatively similar when using an equally-weighted aggregation scheme.} 

\subsection{Tone measurement for earnings press releases and earnings calls}

We compute the tone of earnings press releases and earnings calls in two ways. First, we use the lexicon by \citet[hereafter LM]{LoughranMcdonald2011} as in \citet{huang2013tone}. Second, we use the same GWP model as the one used to compute the tone of media news.\footnote{A third approach is to use two GWP models: one trained on earnings press releases and one on earnings call transcripts. This methodology has the natural advantage of linking directly the words in the corporate disclosures and call transcripts to the market returns. The pitfall with this approach is the low frequency of the publications leading to a relatively low number of texts that we can use for the model-fitting compared to media news articles. The optimal choice of the tone estimation method is case-specific. In the Internet Appendix, Section D, we report the results when computing the tone using GWP models trained on earnings press releases and earnings calls. We find that, for our corpus, using the GWP on media news leads to higher numerical stability and better out-of-sample performance.} The first approach is standard to compute the tone of corporate disclosures in accounting and finance. The second is better aligned with our definition 
of tone as the expected stock return implied by the textual document. The LM and GWP methods applied to earnings press releases and earnings calls allow us to construct four variables to control for the information disclosed by the firm to isolate information provided by the media. We call these variables $\textit{EPRLM}$, $\textit{EPRGWP}$, $\textit{ECLM}$, and $\textit{ECGWP}$.

\section{Media coverage and abnormal tone around earnings announcements}
\label{sec:catdriver}

Earnings announcements provide essential information to investors \citep{basu2013important}. The media provide substantial coverage of earnings announcements, which suggests they play a critical role in distributing information about announcements \citep{tetlock2008more}. In this section, we aim to understand the media environment around earnings announcements better. First, we analyze media coverage near earnings announcements. Second, we look at how media news articles differ from earnings press releases and earnings calls in terms of the topics covered. Third, we analyze the most influential words for the tone obtained with the GWP model when applied to news articles, earnings press releases, and earnings calls. Finally, we analyze the drivers of CAT at and after an announcement using graphical and panel regression analyses.

Our data set contains 23,490 earnings announcements for 598 firms. However, several conditions must be met for an event to be included in our analysis. First, we require that every event-firm observation has a daily tone measure available on the event date, or one day before or one day after. Second, to mitigate potential issues related to missing observations for the estimation of the normal tone model, we require at least ten daily tone measures in the estimation window. Finally, we also require that we have an estimate of the tone for the 
earnings press release and the earnings call. In total, these filters reduce the number of events to 6,394.

We consider an estimation window of 30 days with  an offset of 5 days (i.e., $L = 30$, $K = 5$; the estimation window is $[t_i - 35, t_i - 6]$). We focus our analyses on four post-estimation windows: (i) the pre-event window (i.e., $[t_i - 5, t_i - 2]$); (ii) the event window (i.e., $[t_i - 1, t_i + 1]$); (iii) the short-term post-event window (i.e., $[t_i + 2, t_i + 5]$); and (iv) the long-term post-event 
window (i.e., $[t_i + 2, t_i + 20]$). This setup allows us to consider the pre- and post-event dynamics of the abnormal tone, as well as the reaction at the time of the event.\footnote{The total length of the estimation and event windows is 61 days, which is the typical length of time between a firm's quarterly earnings announcements. This ensures we have non-overlapping events when considering a single firm.}

\subsection{Number of news articles}
\label{sec:distmedium}

We first analyze the media's publication of firm-related news articles around earnings announcements. We expect a spike in the number of articles released during the event window for all three types of media sources. We also expect heterogeneity in the extent of the spike across the types of sources: Newswires are the primary source of news about firms, whereas newspapers and web publications can decide whether to report an event. We thus expect more articles from newswires than from newspapers and web publications.

\begin{figure}[H]
\centering
\singlespacing
\caption{\textbf{Average number of articles around the event date}\\
This figure shows the (cross-section) average number of articles relative to the event date by source types 
across  6,394 quarterly earnings announcement events. Day 0 is the event day, and the gray area denotes the event window.}
\includegraphics[width=\textwidth]{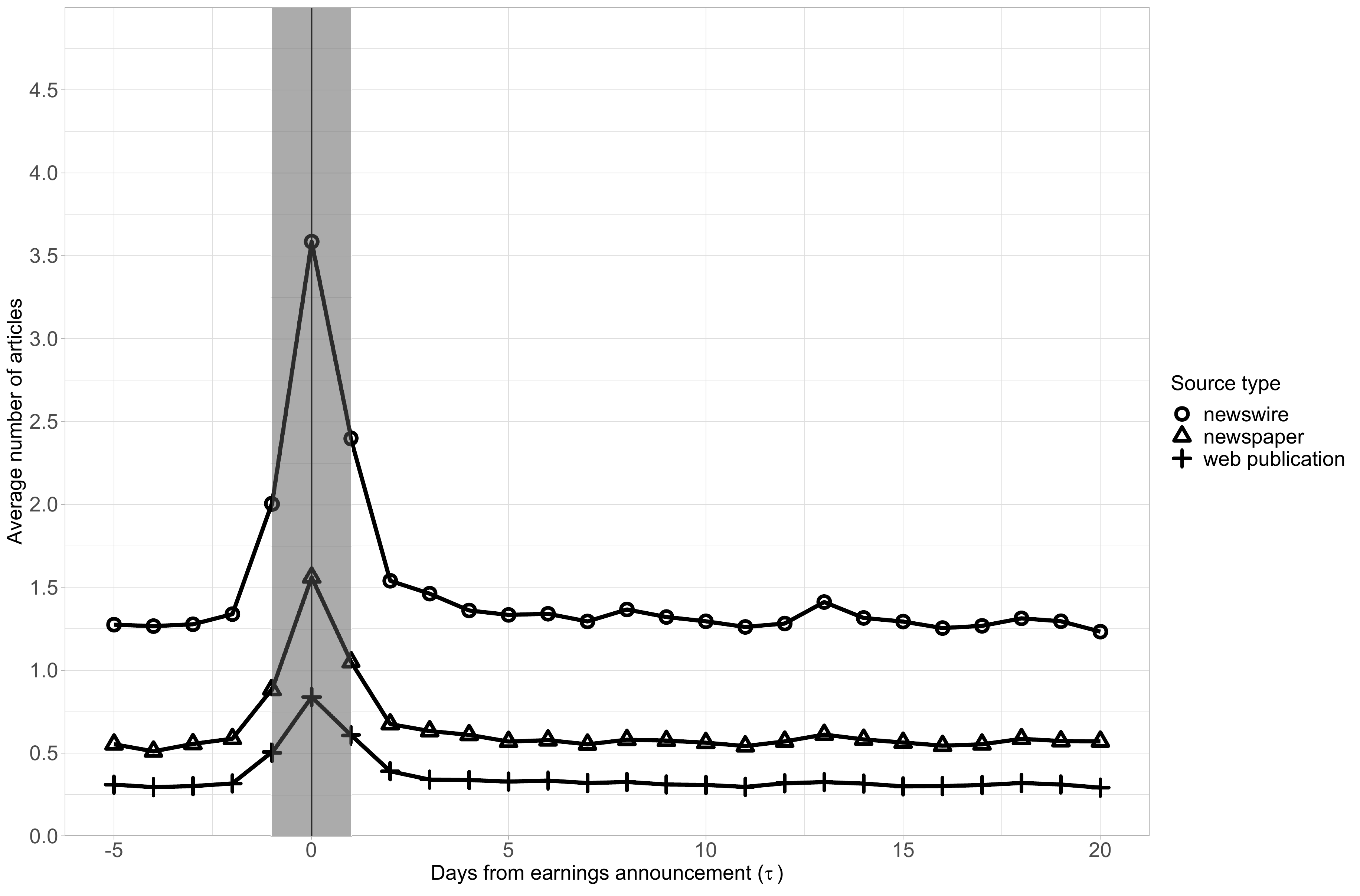}
\label{fig:averagedoctype}
\end{figure}

In Figure~\ref{fig:averagedoctype}, we display the daily average number of articles published around the event date for the three media types. For newswires, newspapers, and web publications, the average number of articles during the event window are $2.66$, $1.16$, and $0.65$, respectively, compared to averages on non-event days of $1.32$, $0.57$, and $0.31$, respectively. While this is in line with our expectations, we also note a substantial number of articles released in the short-term post-event window. We posit that some media outlets are late in reporting about earnings announcements \citep{tetlock2011all}. These articles may contain further analysis or updated information, such as analyst comments on the financial results, on the days following an announcement. 

\insertfloat{Figure~\ref{fig:averagedoctype}}

\subsection{Coverage of topics}

We study the incremental information value of the media by analyzing the difference in content and tone between the newspaper articles and the earnings press releases and earnings calls. The topic analysis is arduous because the class of documents can affect word frequencies across topic and topic prevalence. We address that challenge by using the structural topic model (STM) of \citet{RobertsEtAl2016} that allows us to incorporate the effects of metadata (here a variable indicating whether the text is classified as a news article, an earnings press release, or an earnings call transcript) into the topic model. Before applying STM, we process our corpus of texts following \citet[p.~218]{hansen2018transparency} to identify two-word and three-word collocations, which are sequences of words that have a specific meaning. We estimate the STM with 30 topics.\footnote{Using the approach of \citet{mimno2014low}, we find that the optimal number of topics is 54. \citet{hansen2018transparency} estimate the optimal number of topics at 70, but reduce this to 40 for ease of interpretation. We follow their reasoning.}

In Table~\ref{tab:TopicDifferences}, we report the ten largest (absolute) differences in topic prevalence measured between the corpus of news articles and the earnings press releases (Panel A) and earnings calls (Panel B). We manually assigned a label for each topic based on the most common words or collocations. First, we see that news media are more likely to cover ``analysts,'' ``legal,'' ``innovation,'' and ``stock research'' topics, compared to earnings press releases and earnings calls. The focus on analysts suggests that a specific role of the media is in providing context to the firm results by comparing them against analyst recommendations. The higher prevalence of legal, innovation, and stock research topics points to the information that usually appears in earnings press releases and earnings calls but is highlighted by the media. It is also the case for the topic ``board'' in earnings calls. On the other hand, the themes ``accounting'' and ``earnings'' are less prevalent in news articles than in earnings press releases. The same holds for ``cash/debt management'' in the case of earnings calls. These represent three types of information that the media could have picked up, but in practice, are generally not. Topics such as ``entertainment,'' ``telecommunication,'' ``electricity utility,'' ``pulp and paper,'' and ``consumer goods'' are related to sectors. Finally, note that for earnings call transcripts, more prevalent collocations are typically used to report the discussion of the call (i.e., labeled ``ec-related terms''). This topical analysis illustrates that media focus on certain discussion subjects to contextualize, emphasize, or mitigate specific results compared to earnings press releases and earnings calls.

\insertfloat{Table~\ref{tab:TopicDifferences}}

In addition to comparing the textual content, it is also interesting to compare the words driving the tone in news articles, earnings press releases, and earnings calls around earnings announcements. We provide such an analysis in the Internet Appendix, Section~E. We also summarize the context surrounding the appearance of the important words.

\subsection{Average CAT by level of SUE}

Our previous analyses reveal that the media report substantially about the firm around earnings announcements. 
Moreover, we find variations in the topics covered, the ranking of words driving the tone, and the context 
in which they appear, between news articles and the earnings press releases and earnings calls. Despite these differences, we still expect a positive relation between the tone and the earnings surprise. To test this, we first look at the evolution of the (cross-section) average CAT conditional on different standardized unexpected earnings (SUE) levels. We split the events into five buckets, where the cutoff points for SUE are based on their quintiles. 

\begin{figure}[H]
\centering
\singlespacing
\caption{\textbf{Evolution of the CAT around the event date}\\
This figure shows the evolution of the (cross-section) average CAT for five SUE buckets over the 6,394 quarterly earnings announcements. The buckets are based on the quintiles of SUE from the lowest (\#1) to the highest (\#5). Day 0 is the event day, and the gray area denotes the event window.}
\includegraphics[width=\textwidth]{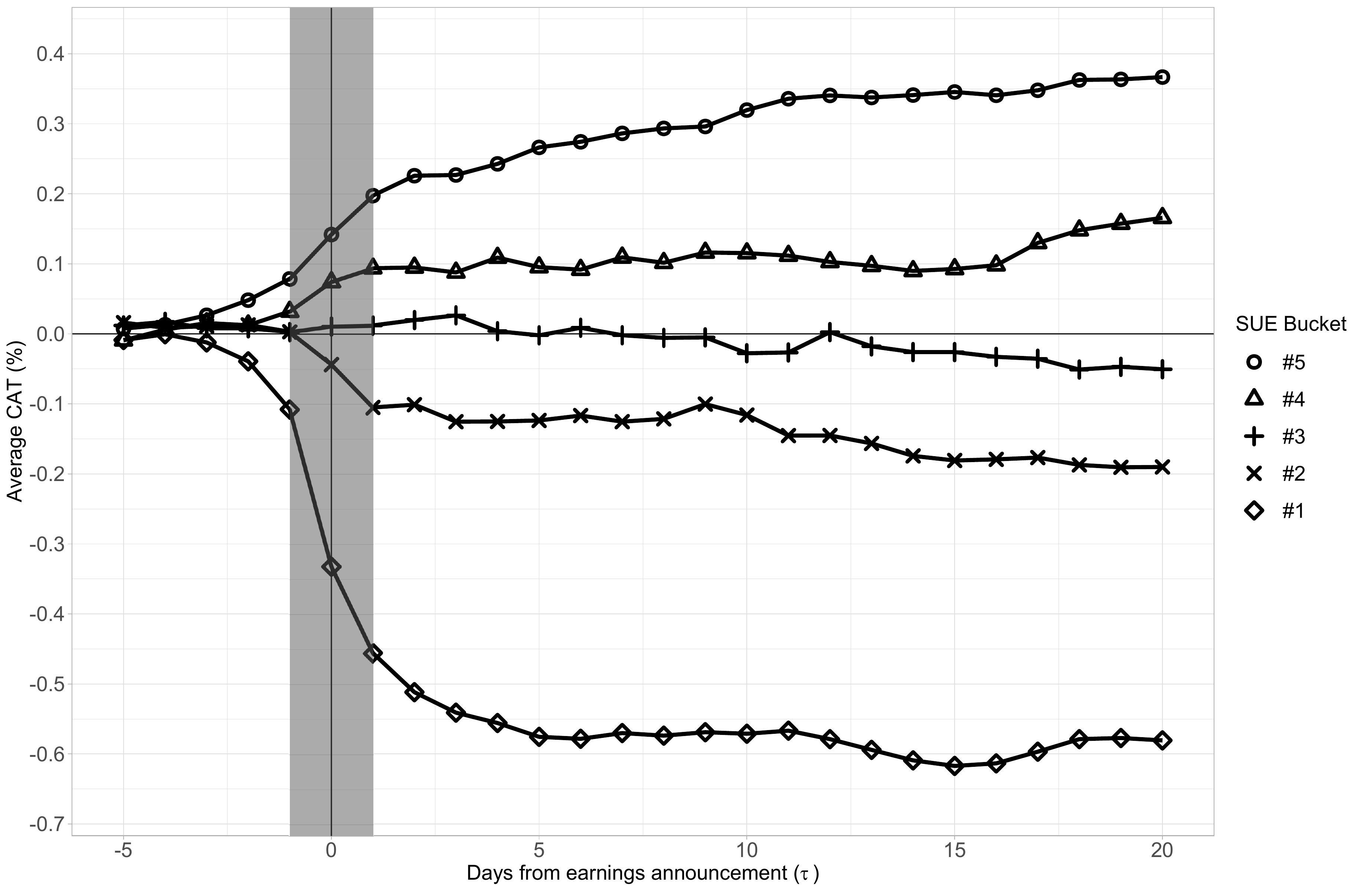}
\label{fig:cat}
\end{figure}

In Figure~\ref{fig:cat}, we display the evolution of the average CAT around the event. During the pre-event window, there is no pattern suggesting that CAT can help predict SUE. While the media reaction appears to start the day before an announcement, this effect is due to overnight earnings announcements. However, during the event window, we find a clear relation between the level of the earnings surprise and the CAT: the media reacts more strongly to negative than positive earnings surprises. For the short-term post-event window, we see a small drift across the negative SUE buckets and the most positive SUE bucket. For the long-term post-event window, we do not see any clear pattern across the buckets. 

\insertfloat{Figure~\ref{fig:cat}}

\subsection{Earnings information in CAT}
\label{ss:cat}

The graphical analysis in Figure~\ref{fig:cat} indicates that the average CAT is related to the level of earnings surprises. We now use a panel regression to study the predictive value of the information available in earnings press releases and earnings calls for the media's abnormal tone about a firm. We focus on three time horizons for the CAT dependent variable: (i) at the event with \CAT{-1}{1}; (ii) the short-term post-event with \CAT{2}{5}; and (iii) the long-term post-event with \CAT{2}{20} window.\footnote{For readability, we remove the event-firm index $i$ for the dependent and explanatory variables.} For the explanatory variables, we consider several variables that journalists could report on or be affected by. We consider  the earnings surprise, $\textit{SUE}$; the negative earnings surprise indicator, $I[\textit{SUE}\!<\!0]$; and their interaction, $\textit{SUE} \!\times\! I[\textit{SUE}\!<\!0]$. We also consider variables capturing the tone of the earnings press release and the earnings call, as these are the source material for journalists. 
We use $\textit{EPRLM}$, $\textit{EPRGWP}$, $\textit{ECLM}$, and $\textit{ECGWP}$, where $\textit{EPR}$ refers to the earnings press release, $\textit{EC}$ to the earnings call, $\textit{LM}$ to the lexicon-based measure, and $\textit{GWP}$ to the GWP-based measure. We include CAT and the cumulative abnormal return, $\textit{CAR}$, before and on the event date to control for past media reporting and to determine whether the media tends to focus on past market reaction. To compute CAR, we follow \citet{mackinlay1997event} and calculate abnormal returns using a market model, where the market is defined as in \citet{fama1992cross}.\footnote{The market is the return of a value-weighted portfolio of all CRSP firms incorporated in the United States and listed on the NYSE, AMEX or NASDAQ. Data are available from  Kenneth R. French's website at \url{http://mba.tuck.dartmouth.edu/pages/faculty/ken.french/data_library.html}.} We include as additional control variables the return on assets, $\textit{ROA}$; the logarithmic book-to-market ratio, $\textit{log(B/M)}$; the logarithmic market value, $\textit{log(M)}$; and firm and year-quarter fixed effects. Finally, we winzorize the variables $\textit{SUE}$, $\textit{ROA}$, $\textit{log(M)}$, and $\textit{log(B/M)}$ at the 1\% and 99\% levels.

Regression results are reported in Table~\ref{tab:catdrivers}. We find a significant relation between the media's abnormal tone and the tone in earnings press releases and earnings calls during the event window. The media's abnormal tone tends to be lower (higher) when the tone of the earnings press release or the earnings call is negative (positive). The latter relation is expected, as the earnings press release and the earnings call are major sources for journalists when writing news articles about the firm's earnings announcement. We also see that the media's abnormal tone tends to be lower when the surprise is negative. This finding is in line with several studies which have identified that the coverage of bad news is disproportionately negative \citep[e.g.,][]{soroka2006good,garz2013unemployment,arango2014bad}. In addition, we find that the media's abnormal tone is positively correlated with its past value. This serial correlation can be attributed to the media's tendency to add context to articles by referring to past information \citep{fink2014rise,mcintyre2018contextualist}. We observe a similar media tone momentum for the short- and long-term post-event CAT reported in the last two columns of Table~\ref{tab:catdrivers}.

Finally, consistent with the view that the media only reports about firms when there is new information, we find that the media's abnormal tone in the long-term post-event window is negatively related to the tone of the earnings press release and the tone of the earnings call. This is in line with the finding of \citet{graffin2016ready} that firms tend to release unrelated positive news in the days immediately surrounding a potentially negatively received acquisition announcement. They suggest that managers may wish to engage in this sort of reputation management in other contexts that can be easily anticipated, such as negative earnings announcement events.

\insertfloat{Table~\ref{tab:catdrivers}}

\section{Media tone and stock market reaction around earnings announcements}
\label{sec:catcar}

We now investigate how news articles relate to movements in stock prices and trading volumes. The attribute of interest is the RCAT of media articles, the residual of the panel regression estimated in the previous section. Its construction is analogous to the one of the abnormal tone of earnings press release analyzed in \citet{huang2013tone} and \citet{arslan2016managers}, who use a similar regression to extract the managerial tone inflation in earnings press releases. In media news analysis, the RCAT  corresponds to the unexpected media tone and thus reflects the effect of transformation of information by the media rather than dissemination. We denote the RCAT over the event window as \RCAT{-1}{1}.

If the media's transformation of the primary information has some value to investors, \RCAT{-1}{1} should be positively correlated with the stock price reaction in the three days around an event. Such a positive reaction would be consistent with the price reaction to the (abnormal) tone in earnings press releases \citep{demers2008soft,Henry2008,davis2012beyond,huang2013tone} and earnings calls \citep{PriceEtAl2012}. It is also consistent with \citet{engelberg2008costly}, who observes a positive correlation between stock returns and the tone of articles from newswires at earnings announcements. In the case where  \RCAT{-1}{1} is a proxy for investor sentiment and the behavior of noise traders, a reversal in stock price should be observed \citep[e.g.,][]{de1990noise,Tetlock2007, dougal2012journalists,garcia2013sentiment}. \citet{huang2013tone} and \citet{arslan2016managers} observe this pattern on the inflated part of the tone of earnings press releases.  

We first study the relation between \RCAT{-1}{1} and the stock market using a graphical analysis of the average CAR based on buckets conditional on \RCAT{-1}{1}. We then use a panel regression to test for the incremental information of the media's abnormal tone compared to the quantitative and qualitative information available in the earnings press release and earnings call. We also study whether the type of news matters and how \RCAT{-1}{1} relates to the abnormal volume around the earnings announcement. To avoid any look-ahead bias, we estimate the panel regression up to and including the year-quarter in which \CAT{-1}{1} is present to compute \RCAT{-1}{1}. Thus, our analysis relies on data from 2004:Q1 to 2016:Q4.

\subsection{Average CAR by level of RCAT}
\label{sec:carrcat}

We first tackle this question from an exploratory data analysis viewpoint and investigate the evolution of the (cross-section) average CAR conditional on the level of \RCAT{-1}{1} and the direction of the earnings surprise. Specifically, we build six buckets based on \RCAT{-1}{1} terciles conditional on the sign of the earnings surprise (three for positive earnings surprises and three for negative). 

\begin{figure}[H]
\centering
\singlespacing
\caption{\textbf{Evolution of the CAR around the event date}\\
This figure shows the evolution of the (cross-section) average CAR for six RCAT buckets (three for positive earnings surprises and three for negative earnings surprises) over 4,662 quarterly earnings announcements. The buckets are based on terciles of \RCAT{-1}{1} from the lowest (\#1) to the highest (\#3) conditional on the sign of the corresponding value of SUE. Day 0 is the event day, and the gray area denotes the event window.}
\includegraphics[width=\textwidth]{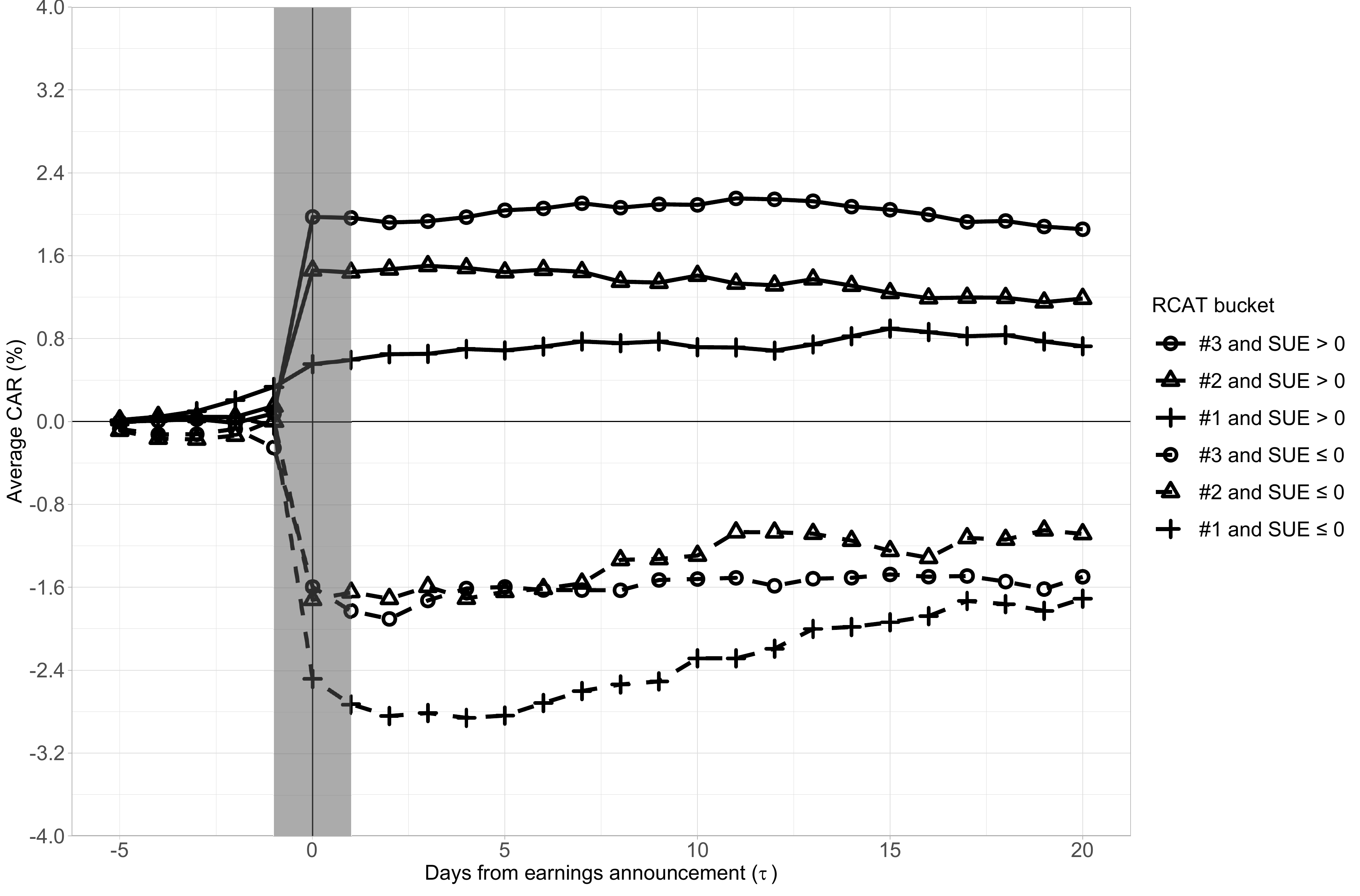}
\label{fig:car}
\end{figure}

In Figure~\ref{fig:car}, we display the evolution of the average CAR around the event. First, even though \RCAT{--1}{1} is orthogonal to the information that can be derived from earnings press releases and earnings calls, we find a sizeable positive relation between the abnormal return and RCAT during the event window. The relation appears to be stronger when an earnings surprise is negative.  Second, in the long-term post-event window, we observe a reversal in stock price when an earnings announcement is negative. This reversal appears to be stronger the lower \RCAT{-1}{1} is, and it is partial, at least in the 20 trading days after the event. We do not observe a reversal for positive earnings announcements.

\insertfloat{Figure~\ref{fig:car}}

\subsection{Stock price reaction information in RCAT}\label{main}

We now perform more extensive analyses using panel regressions to assess the relation between stock returns and the media's transformation of primary information while controlling for confounding factors. We are interested in both the immediate and delayed responses of the CAR to RCAT.  Consistent with our previous analyses, we focus on three time horizons for the CAR dependent variable: (i) at the event with \CAR{-1}{1}; (ii) the short-term post-event window with \CAR{2}{5}; and (iii) long-term post-event window with \CAR{2}{20}. We include the same control variables as in the CAT regressions in Section~\ref{sec:catdriver}.

We report the results of the regressions in Table~\ref{tab:regcontrolwp}. First, we find a positive and significant relation between RCAT and CAR in the event window. An increase of one standard deviation in \RCAT{-1}{1} is associated with an abnormal return of 0.84\% (i.e., the coefficient of 0.825 times the standard deviation of \RCAT{-1}{1} that is 1.018) at the time of the event. Second, although we do not find that \RCAT{-1}{1} has a significant relation with \CAR{2}{5}, we find a negative and significant relation with \CAR{2}{20}. In this case, an increase of one standard deviation in \RCAT{-1}{1} is associated with an abnormal return of -0.36\% in the next 19 trading days. Thus, we observe a partial reversal of the initial reaction. These results suggest that \RCAT{-1}{1} conveys information about the market's response around an earnings announcement. This information is incremental to the qualitative and quantitative information available in the earnings press release and the earnings call, and investors overreact to this information.

\insertfloat{Table~\ref{tab:regcontrolwp}}

The above result is consistent with the hypothesis that the abnormal tone attracts the attention of investors, leading to an immediate overreaction and a subsequent price 
reversal \citep{da2011search}. The result contradicts the underreaction hypothesis of \citet{engelberg2008costly}. 

\subsection{ Stock price information in RCAT conditional on media source}
\label{sec:regcatcarsource}

Newswires have different business models and target audiences than newspapers and web publications. The former are typically available to professional investors on platforms such as the Bloomberg terminal, whereas newspapers and web publications have a broader reach and are often consulted by retail investors. Hence, it is possible that the relation between RCAT and CAR is different for newswires than for newspapers and web publications. To test this, we aggregate the $\textit{RATC}_{d,i,\tau}\mbox{\footnotesize{(-1,1)}}$ by two groups: one for newswires, and one for newspapers and web publications. We refer to these values as \NWRCAT{-1}{1} for the newswires and \NPRCAT{-1}{1} for newspapers and web publications.

In Table~\ref{tab:contregression}, we report the results of regressions that separate the effect of RCAT for the different media sources. We find a stock price reaction to RCAT for both groups at the event: an increase of one standard deviation in \RCAT{-1}{1} for newswires is associated with an abnormal return of 0.56\% (0.53\% for newspapers and web publications). Interestingly, we find a partial reversal for all media sources. This reversal appears to be larger for newswires than for newspapers and web publications, even though the immediate reaction is economically larger for newspapers. However, at each time horizon, the difference is not significantly different from zero at standard significance levels.\footnote{The Wald test statistics of equal coefficients 
is 0.86 ($p$-value = 0.30) for \CAR{-1}{1}, 1.33 ($p$-value = 0.18) for \CAR{2}{5}, and 0.85 ($p$-value = 0.40) for \CAR{2}{20}.}

\insertfloat{Table~\ref{tab:contregression}}

\subsection{Abnormal turnover and RCAT}

The partial reversal in stock prices points to an overreaction by investors to the incremental information provided by the media. \citet{hou2009tale} argue that one of the necessary conditions for overreaction is attention, which can be proxied by trading volumes \citep{barber2008all}. We thus investigate whether the magnitude of the incremental information conveyed by RCAT (absolute value of RCAT) has any relation to trading volume around earnings announcements. We use the abnormal logarithmic share turnover, where we compute the abnormal logarithmic share turnover of the event window using the mean-adjusted model of \citet{campbell1996measuring}. Taking the cumulative sum of the abnormal logarithmic share turnovers over relative times leads to the CAST measure.

\begin{figure}[H]
\centering
\singlespacing
\caption{\textbf{Evolution of the CAST around the event date}\\
This figure shows the evolution of the (cross-section) average CAST for six RCAT buckets (three for positive earnings surprises and three for negative earnings surprises), where the RCAT buckets are based on terciles of |\RCAT{-1}{1}| from the lowest (\#1) to the highest (\#3)) over 4,662 quarterly earnings announcement events. Day 0 is the event day, and the gray area denotes the event window.}
\includegraphics[width=\textwidth]{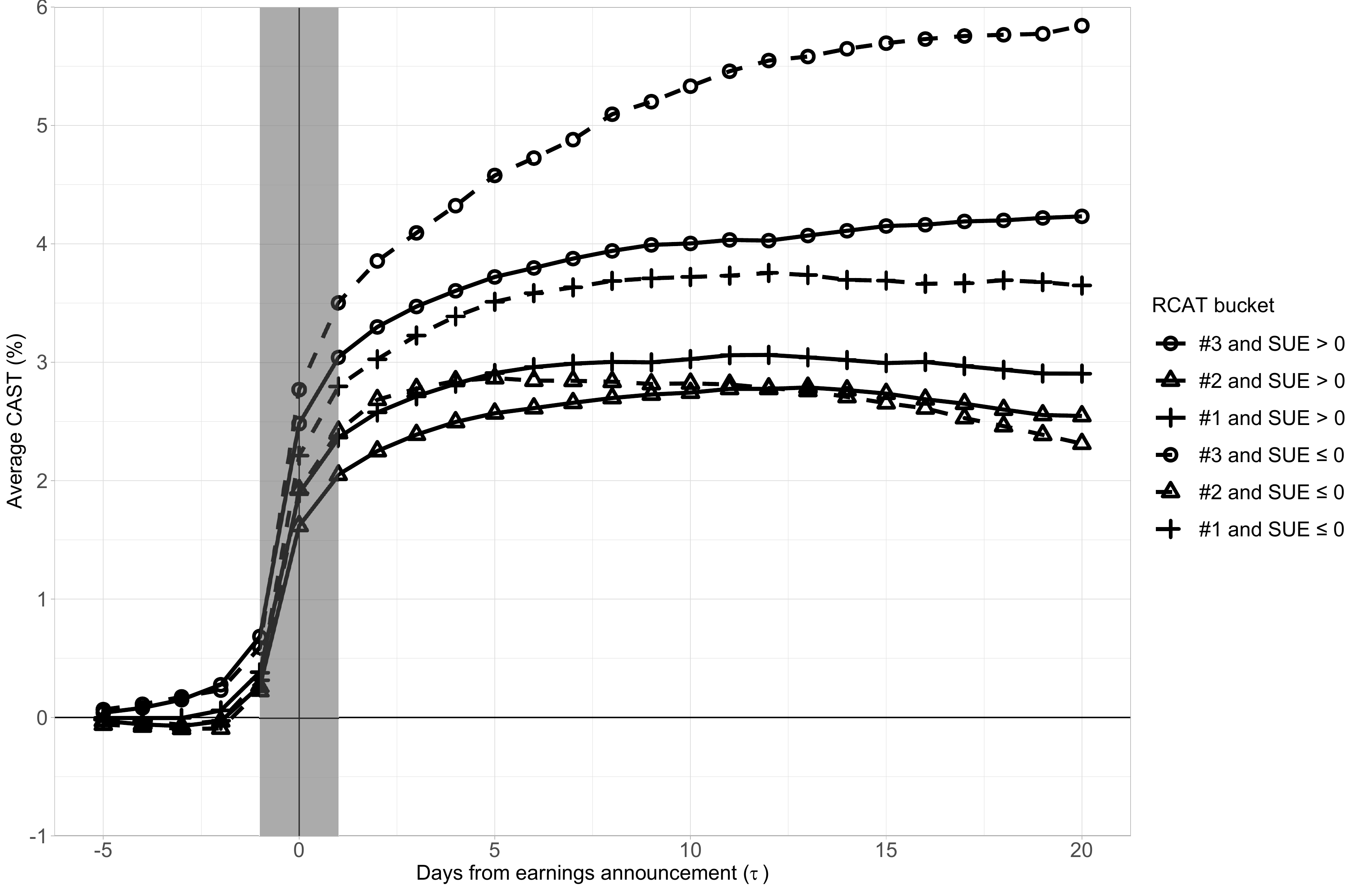}
\label{fig:calvt}
\end{figure}

In Figure~\ref{fig:calvt}, we display the evolution of the (cross-sectional) average CAST conditioned on the magnitude of RCAT and the sign of SUE. For all buckets, the CAST increases at the event date. The increase is the most pronounced when the magnitude of RCAT is the highest, thus supporting our conjecture that investors overreact to incremental information conveyed by RCAT.

\insertfloat{Figure~\ref{fig:calvt}}

In Table~\ref{tab:regcalvt}, we report results of the regressions for \CAST{-1}{1}. As explanatory variables, we consider the magnitude of \RCAT{-1}{1}, as well as controls used in our previous analyses. We find a significant positive contemporaneous relation between the magnitude of \RCAT{-1}{1} and the abnormal volume, again supporting an overreaction. When looking at different media sources, we find that both the incremental information extracted from newswires or newspapers and web publications has a positive and significant relation with trading volume.\footnote{The Wald test statistics of equal coefficients is 0.42 ($p$-value = 0.67).} 

\section{Conclusion}
\label{sec:conclusion}

The news media plays a central role in disseminating information about corporate earnings. While one could try to find anecdotal evidence of news articles that help to explain stock price reactions around earnings announcements, the scientific challenge is to formalize this exercise into a framework that aggregates the information from news articles into statistics that represent the incremental information that can be distilled from information disclsed by the firms, the earnings press releases, and the earnings calls.

We conduct a tone-based event study to quantify the added qualitative information provided in news articles about earnings announcements. We apply our novel methodology to S\&P 500 firms from 2000 to 2016. We focus on newswires, newspapers, and web publications.

We show that the abnormal tone in media articles helps explain price reactions around an earnings announcement and predicts subsequent price dynamics in the 20 days following the event. While there is a significant positive contemporaneous association between abnormal tone and price movement, the abnormal tone is negatively associated with post-event price movement. We conclude that the media provides additional information not included in the information from earnings press releases and earnings calls, and on aggregate, market participants overreact to it. Complementary analyses of the traded volumes support these findings.

\newpage
\singlespacing
\bibliographystyle{elsarticle-harv}

\begingroup
\setlength{\bibsep}{10pt}
\setstretch{1}

\endgroup

\newpage
\begin{table}[H]
\centering
\singlespacing
\caption{\textbf{Number of articles and coverage}\\ This table reports the average, minimum, and maximum number of news articles, as well as the daily percentage of news media coverage, for the 598 non-financial historical constituents of the S\&P 500 Index over the 2000:Q1 to 2016:Q4 period. We split the constituents into four buckets, where the 1st (4th) bucket is composed of the firms with the lowest (highest) average market-capitalization value when they were constituents of the index.}
\label{tab:coverage}
\scalebox{0.95}{
\begin{tabular}{lccccccc}
\toprule
\multicolumn{8}{l}{Panel A: All sources} \\
\midrule
&& \multicolumn{3}{c}{News articles} & \multicolumn{3}{c}{News coverage}\\
Bucket & \multicolumn{1}{c}{Number of firms} 
& \multicolumn{1}{c}{Average} & \multicolumn{1}{c}{Min} & \multicolumn{1}{c}{Max}  
& \multicolumn{1}{c}{Average} & \multicolumn{1}{c}{Min} & \multicolumn{1}{c}{Max}\\
\midrule
\#1 & 150 & 948 & 4 & 12,496 & 20.67 & 0.76 & 85.26 \\ 
\#2 & 149 & 1,447 & 2 & 11,096 & 28.23 & 0.27 & 81.04 \\ 
\#3 & 149 & 2,645 & 23 & 19,716 & 35.65 & 1.65 & 90.97 \\ 
\#4 & 150 & 10,449 & 20 & 86,560 & 54.82 & 4.32 & 100.00 \\ 
All & 598 & 3,878 & 2 & 86,560 & 34.85 & 0.27 & 100.00 \\[8pt]  
\midrule
\multicolumn{8}{l}{Panel B: Newswires} \\ 
\midrule
&& \multicolumn{3}{c}{News articles} & \multicolumn{3}{c}{News coverage}\\
Bucket & \multicolumn{1}{c}{Number of firms} 
& \multicolumn{1}{c}{Average} & \multicolumn{1}{c}{Min} & \multicolumn{1}{c}{Max}  
& \multicolumn{1}{c}{Average} & \multicolumn{1}{c}{Min} & \multicolumn{1}{c}{Max}\\
\midrule
\#1 & 150 & 680 & 1 & 7,415 & 17.04 & 0.32 & 71.40 \\ 
\#2 & 149 & 1,035 & 2 & 8,450 & 22.84 & 0.27 & 73.55 \\ 
\#3 & 149 & 1,817 & 7 & 12,325 & 28.45 & 1.32 & 80.69 \\ 
\#4 & 150 & 5,715 & 17 & 45,072 & 45.75 & 3.02 & 98.98 \\ 
All & 598 & 2,315 & 1 & 45,072 & 28.53 & 0.27 & 98.98 \\[8pt] 
\midrule
\multicolumn{8}{l}{Panel C: Newspapers} \\ 
\midrule
&& \multicolumn{3}{c}{News articles} & \multicolumn{3}{c}{News coverage}\\
Bucket & \multicolumn{1}{c}{Number of firms} 
& \multicolumn{1}{c}{Average} & \multicolumn{1}{c}{Min} & \multicolumn{1}{c}{Max}  
& \multicolumn{1}{c}{Average} & \multicolumn{1}{c}{Min} & \multicolumn{1}{c}{Max}\\
\midrule
\#1 & 150 & 163 & 0 & 2,427 & 4.81 & 0.00 & 46.63 \\ 
\#2 & 149 & 211 & 0 & 2,256 & 5.68 & 0.00 & 56.79 \\ 
\#3 & 149 & 477 & 0 & 5,044 & 9.53 & 0.00 & 58.53 \\ 
\#4 & 150 & 2,874 & 3 & 34,459 & 26.91 & 0.35 & 99.87 \\ 
All & 598 & 933 & 0 & 34,459 & 11.75 & 0.00 & 99.87 \\[8pt]
\midrule
\multicolumn{8}{l}{Panel D: Web publications} \\
\midrule
&& \multicolumn{3}{c}{News articles} & \multicolumn{3}{c}{News coverage}\\
Bucket & \multicolumn{1}{c}{Number of firms} 
& \multicolumn{1}{c}{Average} & \multicolumn{1}{c}{Min} & \multicolumn{1}{c}{Max}  
& \multicolumn{1}{c}{Average} & \multicolumn{1}{c}{Min} & \multicolumn{1}{c}{Max}\\
\midrule
\#1 & 150 & 104 & 0 & 2,654 & 3.14 & 0.00 & 42.81 \\ 
\#2 & 149 & 199 & 0 & 2,633 & 5.84 & 0.00 & 40.43 \\ 
\#3 & 149 & 351 & 0 & 2,347 & 8.40 & 0.00 & 58.97 \\ 
\#4 & 150 & 1,858 & 0 & 31,024 & 20.18 & 0.00 & 99.87 \\ 
All & 598 & 629 & 0 & 31,024 & 9.40 & 0.00 & 99.87 \\ 
\bottomrule
\end{tabular}}
\end{table}

\newpage
\begin{sidewaystable}[H] 
\centering 
\caption{\textbf{Difference in topic prevalence}\\
This table reports topics with the largest absolute difference in prevalence between media news articles published near earnings announcements and earnings press releases (Panel A) or earnings calls (Panel B). Column $\Delta P$ reports the topic prevalence's difference between news articles and earning press releases or earnings calls together with its 99\% confidence interval in squared brackets. A positive (negative) value indicates that the topic is more (less) prevalent in news articles than earnings press releases or earnings calls. Column ``Topic'' reports the topic name (manually assigned). The last column reports the five most common words or collocations (sequence of words with a specific meaning) with a high probability of occurring in the topic. Note that stop words (e.g., ``and'', ``the'', ``to'', etc.) have been removed.}
\label{tab:TopicDifferences}
\scalebox{0.85}{
\begin{tabular}{ccl}
\toprule
\multicolumn{3}{l}{Panel A: News articles vs. earnings press releases}\\
\midrule
$\Delta P$   & Topic & High-probability word or collocation\\
\midrule
0.112 [0.108;0.117]& analysts & analysts, chief\_executive, consensus\_revenue\_estimate, consensus\_estimate, profit\_per\_share\\
0.083 [0.080;0.087]& legal & court, investigation, lawsuit, attorney, bankruptcy           \\
0.059 [0.056;0.063] & innovation & award, technology, leader, developer, manager              \\
0.056 [0.052;0.060]& stock research & investment\_research, sentiment, price\_target, economist, rating \\
0.022 [0.019;0.024]& entertainment & show, radio, entertainment, football, music  \\
0.019 [0.016;0.022]& telecommunication & connection, telecommunication, phone, carrier, network \\
-0.013 [-0.017;0.009]& electricity utility & megawatt, electricity, outage, load, grid\\
-0.022 [-0.025;0.020]& pulp and paper & pulp, tissue, paper, wood, raw\_material \\
-0.092 [-0.095;-0.089]  & accounting& deferred\_income, total\_current\_liabilities, net\_cash, amortization\_of\_goodwill, provision\_for\_income \\
-0.234 [-0.238;-0.230]  & earnings & news\_release, operating\_performance, foreign\_currency\_exchange, other\_risks, non-gaap\_financial\_measure\\[0.3cm]
\midrule
\multicolumn{3}{l}{Panel B: News articles vs. earnings calls}\\
\midrule
$\Delta P$   & Topic & High-probability word or collocation\\
\midrule
0.119 [0.116;0.123] & analysts & analysts, chief\_executive, consensus\_revenue\_estimate, consensus\_estimate, profit\_per\_share \\
0.080 [0.077;0.083] & legal & court, investigation, lawsuit, attorney, bankruptcy           \\
0.075 [0.072;0.078] & innovation & award, technology, leader, developer, manager              \\
0.070 [0.066;0.073]& stock research & investment\_research, sentiment, price\_target, economist, rating               \\
0.032 [0.020;0.036]& earnings  & operating\_performance, foreign\_currency\_exchange, other\_risks, non-gaap\_financial\_measure, news\_release\\
0.021 [0.019;0.023]& board & director, shareholder,  committee, holder, annual meeting\\
0.018 [0.016;0.020]& telecommunication & connection, telecommunication, phone, carrier, network \\
-0.021 [-0.024;-0.019]& consumer goods & merchandise, footwear, jewelry, shopping, fashion \\
-0.063 [-0.065;-0.061] & cash/debt management & receivables, liquidity, working\_capital, debt\_reduction, cap\_ex               \\
-0.261 [-0.263;-0.258] & ec-related terms & analyst\_presentation\_operator, transcription, speakers, follow\_up, question               \\
\bottomrule
\end{tabular}}
\end{sidewaystable} 

\newpage
\begin{table}[H] 
\centering 
\caption{\textbf{CAT regression results}\\
This table reports the panel regression results of the CAT computed over various time spans on firm- and event-specific variables. Dependent variables are the CAT computed: (i) at the event with \CAT{-1}{1}; (ii) short-term post-event with \CAT{2}{5}; and (iii) long-term post-event with \CAT{2}{20}. See Section~\ref{ss:cat} for a description of the explanatory variables. $^{*}$, $^{**}$, and $^{***}$ denote statistical significance at the 10\%, 5\%, and 1\% levels, respectively. The standard errors are computed using the double-clustered (year-quarter and firm) standard error estimator \citep{petersen2009estimating} and are reported in parentheses below the parameter estimates.}
\label{tab:catdrivers}
\scalebox{1.00}{ 
\begin{tabular}{@{\extracolsep{2pt}}lcccc} 
\toprule
& \CAT{-1}{1} & \CAT{2}{5} & \CAT{2}{20} \\  
\midrule
$\textit{SUE}$ 
& 0.192 & 0.144 & 0.118 \\ 
& (0.126) & (0.103) & (0.243) \\ 
$I[\textit{SUE}\!<\!0]$ 
& -0.003$^{***}$ & 0.00000 & 0.002$^{**}$ \\ 
& (0.0005) & (0.0004) & (0.001) \\ 
$\textit{SUE} \!\times\! I[\textit{SUE}\!<\!0]$ 
& -0.362$^{*}$ & -0.183 & 0.124 \\ 
& (0.220) & (0.134) & (0.286) \\ 
$\textit{EPRLM}$ 
& 0.007 & -0.001 & -0.055 \\ 
& (0.020) & (0.021) & (0.048) \\ 
$\textit{EPRGWP}$ 
& 0.446$^{***}$ & -0.030 & -0.094 \\ 
& (0.058) & (0.048) & (0.127) \\ 
$\textit{ECLM}$ 
& 0.255$^{***}$ & 0.058 & -0.002 \\ 
& (0.042) & (0.045) & (0.098) \\
$\textit{ECGWP}$ 
& 0.480$^{***}$ & -0.063 & -0.361$^{*}$ \\ 
& (0.091)& (0.083) & (0.186)\\ 
\CAT{-5}{-2} 
& 0.223$^{***}$ & 0.203$^{***}$ & 0.639$^{***}$ \\ 
& (0.026) & (0.023)& (0.054) \\ 
\CAT{-1}{1} 
&  & 0.204$^{***}$ & 0.556$^{***}$ \\ 
&  & (0.019)& (0.054)\\ 
\CAR{-5}{-2} 
& 0.001 &  &  \\ 
& (0.005) &  &  \\ 
\CAR{-1}{1} 
&  & 0.001 & 0.003 \\ 
&  & (0.002) & (0.006) \\ 
$\textit{ROA}$ 
& 0.009 & 0.011 & 0.047 \\ 
& (0.011) & (0.010) & (0.030) \\ 
$\textit{log(B/M)}$ 
&-0.002$^{*}$ & 0.002 & 0.003 \\ 
& (0.001) & (0.001) & (0.003) \\ 
$\textit{log(M)}$ 
& 0.0001 & 0.0002 & -0.0004 \\ 
& (0.0004) & (0.0004) & (0.001) \\[.2cm] 
Firm fixed effects&Yes&Yes&Yes\\
Year-quarter fixed effects &Yes&Yes&Yes\\
Observations & 6,394 & 6,394 & 6,394 \\ 
$\text{R}^{2}$ ($\times 100$) & 10.0 & 9.7 & 13.0 \\ 
\bottomrule
\end{tabular}}
\end{table}

\newpage
\begin{table}[H] 
\centering 
\caption{\textbf{CAR regression results}\\ 
This table reports the panel regression results on the relation between \RCAT{-1}{1} and the CAR computed over various time spans. Dependent variables are CAR computed: (i) at the event with \CAR{-1}{1}; (ii) short-term post-event with \CAR{2}{5}; and (iii) long-term post-event with \CAR{2}{20}. See Section~\ref{ss:cat} for a description of the explanatory variables. $^{*}$, $^{**}$, and $^{***}$ denote statistical significance at the 10\%, 5\%, and 1\% levels, respectively. The standard errors are computed using the double-clustered (year-quarter and firm) standard error estimator \citep{petersen2009estimating} and are reported in parentheses below the parameter estimates.}
\label{tab:regcontrolwp} 
\scalebox{1.00}{
\begin{tabular}{@{\extracolsep{2pt}}lccc} 
\toprule
& \multicolumn{1}{c}{\CAR{-1}{1}} & \multicolumn{1}{c}{\CAR{2}{5}} & \multicolumn{1}{c}{\CAR{2}{20}} \\ 
\midrule
\RCAT{-1}{1} 
& 0.825$^{***}$& -0.029 & -0.356$^{***}$\\ 
& (0.109) & (0.055) & (0.120) \\ 
$\textit{SUE}$ 
& 5.991$^{***}$ & 0.108 & -2.541$^{***}$ \\ 
& (0.757)& (0.348) & (0.839)\\ 
$I[\textit{SUE}\!<\!0]$  
& -0.019$^{***}$ & 0.001 & 0.005 \\ 
& (0.003) & (0.001) & (0.003) \\ 
$\textit{SUE} \!\times\! I[\textit{SUE}\!<\!0]$
& -6.635$^{***}$ & 0.337 & 3.219$^{**}$ \\ 
& (0.944) & (0.708) &   (1.314)\\ 
$\textit{EPRLM}$  
& 0.168 & 0.010 & 0.278$^{*}$ \\ 
& (0.109) & (0.062) & (0.155) \\ 
$\textit{EPRGWP}$  
& -0.136 & 0.102 & -0.227 \\  
& (0.249) & (0.130) & (0.422) \\ 
$\textit{ECLM}$  
& 0.924$^{***}$ & -0.005 & -0.703$^{**}$ \\ 
& (0.218) & (0.130) & (0.319) \\ 
$\textit{ECGWP}$  
& 1.665$^{***}$ & -0.928$^{***}$ & -0.621 \\ 
& (0.563) & (0.292) & (0.676) \\ 
\CAR{-5}{-2} 
& 0.089 &  &  \\ 
& (0.060) &  &  \\ 
\CAR{-1}{1} 
&  & 0.059$^{***}$ & 0.192$^{***}$ \\ 
&  & (0.015) & (0.027) \\ 
$\textit{ROA}$ 
& -0.010 & 0.141$^{***}$ & 0.296$^{**}$ \\ 
& (0.086) & (0.051) & (0.130) \\ 
$\textit{log(B/M)}$  
& -0.014$^{*}$ & 0.017$^{***}$ & 0.054$^{***}$ \\ 
& (0.008) & (0.005) & (0.012) \\ 
$\textit{log(M)}$
& 0.001 & -0.007$^{***}$& -0.035$^{***}$\\ 
& (0.004) & (0.003) & (0.006) \\[.2cm]
Firm fixed effects&Yes&Yes&Yes\\
Year-quarter fixed effects &Yes&Yes&Yes\\
Observations & 4,662  & 4,662 & 4,662  \\ 
$\text{R}^{2}$ ($\times 100$) & 12.5 & 2.8 & 6.8 \\
\bottomrule
\end{tabular}}
\end{table} 

\begin{table}[H] 
\centering 
\caption{\textbf{CAR regression results for source-specific RCAT}\\
This table reports the panel regression results on the relation of source-specific \RCAT{-1}{1} on the CAR computed over various time spans. Dependent variables are the CAR computed: (i) at the event with \CAR{-1}{1}; (ii) short-term post-event with \CAR{2}{5}; and (iii) long-term post-event with \CAR{2}{20}. For the explanatory variables, we consider the contribution of newswires, \NWRCAT{-1}{1}; and the contribution of the newspapers and web publications, \NPRCAT{-1}{1}. See Section~\ref{ss:cat} for a description of the other explanatory variables. $^{*}$, $^{**}$, and $^{***}$ denote statistical significance at the 10\%, 5\%, and 1\% levels, respectively. The standard errors are computed using the double-clustered (year-quarter and firm) standard error estimator \citep{petersen2009estimating} and are reported in parentheses below the parameter estimates.}
\label{tab:contregression} 
\scalebox{1.00}{
\begin{tabular}{@{\extracolsep{2pt}}lccc} 
\toprule
& \multicolumn{1}{c}{\CAR{-1}{1}} & \multicolumn{1}{c}{\CAR{2}{5}} & \multicolumn{1}{c}{\CAR{2}{20}} \\ 
\midrule
\NWRCAT{-1}{1}
& 0.785$^{***}$ & -0.076 & -0.497$^{**}$  \\ 
& (0.133) & (0.078) & (0.220) \\ 
\NPRCAT{-1}{1}
& 0.881$^{***}$ & 0.034 & -0.168 \\ 
& (0.165) & (0.078) & (0.163) \\ 
$\textit{SUE}$ 
& 5.986$^{***}$ & 0.104 & -2.554$^{***}$\\ 
& (0.755) & (0.348) & (0.853) \\ 
$I[\textit{SUE}\!<\!0]$ 
& -0.019$^{***}$ & 0.001 & 0.004 \\ 
& (0.003) & (0.001) & (0.003) \\ 
$\textit{SUE} \!\times\! I[\textit{SUE}\!<\!0]$
& -6.623$^{***}$ & 0.348 & 3.252$^{***}$\\ 
& (0.941) & (0.706) & (1.218) \\ 
$\textit{EPRLM}$ 
& 0.169 & 0.011 & 0.281$^{*}$ \\ 
& (0.109) & (0.062) & (0.155) \\ 
$\textit{EPRCC}$ 
& -0.139 & 0.099 & -0.237 \\ 
& (0.249) & (0.131) & (0.390) \\ 
$\textit{ECLM}$ 
& 0.918$^{***}$ & -0.011 & -0.720$^{**}$\\ 
& (0.217) & (0.130) & (0.348)\\ 
$\textit{ECGWP}$ 
& 1.671$^{***}$& -0.921$^{***}$& -0.602 \\ 
& (0.565) & (0.293) & (0.995) \\ 
\CAR{-5}{-2} 
& 0.089 &  &  \\ 
& (0.060) &  &  \\ 
\CAR{-1}{1} 
&  & 0.059$^{***}$ & 0.192$^{***}$\\ 
&  & (0.015) & (0.026) \\ 
$\textit{ROA}$ 
& -0.010 & 0.142$^{***}$ & 0.298$^{**}$ \\ 
& (0.085) & (0.051) & (0.136)\\ 
$\textit{log(B/M)}$ 
& -0.014$^{*}$ & 0.017$^{***}$ & 0.054$^{***}$ \\ 
& (0.008) & (0.005) & (0.013)\\ 
$\textit{log(M)}$ 
& 0.001 & -0.007$^{***}$& -0.035$^{***}$ \\ 
& (0.004) & (0.003) & (0.008) \\[.2cm]
Firm fixed effects&Yes&Yes&Yes\\
Year-quarter fixed effects &Yes&Yes&Yes\\
Observations &  4,662 &  4,662 &  4,662 \\ 
$\text{R}^{2}$ ($\times 100$) & 12.5 & 2.8 & 6.9 \\ 
\bottomrule
\end{tabular}}
\end{table} 

\newpage
\begin{table}[H] \centering 
\caption{\textbf{CAST regression results}\\	
This table reports the panel regression results on the relation between the absolute value of \RCAT{-1}{1} (and group equivalent) and \CAST{-1}{1}. We use a mean-adjusted model for the abnormal share turnover. See Section~\ref{ss:cat} for a description of the explanatory variables. $^{*}$, $^{**}$, and $^{***}$ denote statistical significance at the 10\%, 5\%, and 1\% levels, respectively. The standard errors are computed using the double-clustered (year-quarter and firm) standard error estimator \citep{petersen2009estimating} and are reported in parentheses below the parameter estimates.}
\label{tab:regcalvt}
\scalebox{1.00}{
\begin{tabular}{@{\extracolsep{2pt}}lcc} 
\toprule
&  \multicolumn{2}{c}{\CAST{-1}{1}} \\ 
\midrule
|\RCAT{-1}{1}|
& 0.120$^{**}$ &  \\  
& (0.054) &  \\ 
|\NWRCAT{-1}{1}|
&  & 0.144$^{*}$ \\ 
&  & (0.079) \\ 
|\NPRCAT{-1}{1}| 
&  & 0.196$^{*}$ \\  
&  & (0.103)\\ 
\textit{SUE}
& 0.701$^{**}$ & 0.701$^{**}$ \\ 
& (0.276) & (0.275) \\ 
$I[\textit{SUE}\!<\!0]$
& 0.004$^{***}$ & 0.004$^{***}$ \\ 
& (0.001) & (0.001) \\ 
$\textit{SUE} \!\times\! I[\textit{SUE}\!<\!0]$ 
& -1.383$^{***}$ & -1.386$^{***}$ \\ 
& (0.521) & (0.522) \\ 
$|\textit{EPRLM}\,|$
& -0.135$^{***}$ & -0.133$^{***}$ \\ 
& (0.048) & (0.048) \\ 
$|\textit{EPRGWP}\,|$
& 0.025 & 0.026 \\
& (0.151) & (0.153) \\
$|\textit{ECLM}\,|$
& 0.010 & 0.013 \\ 
& (0.123) & (0.122) \\  
$|\textit{ECGWP}\,|$
& -0.730$^{***}$ & -0.733$^{***}$ \\
& (0.243) & (0.241) \\ 
\CAST{-5}{-2}  
& 0.574$^{***}$& 0.574$^{***}$ \\ 
& (0.051) & (0.052) \\ 
|\CAR{-5}{-2}|
& -0.027 & -0.027 \\ 
& (0.021) & (0.021) \\  
|\textit{ROA}|
& 0.033 & 0.034 \\ 
& (0.040) & (0.040) \\ 
$\textit{log(B/M)}$
& -0.007$^{**}$ & -0.007$^{**}$ \\ 
& (0.003) & (0.003) \\ 
$\textit{log(M)}$ 
& -0.012$^{***}$ & -0.012$^{***}$ \\ 
& (0.002) & (0.002)  \\[.2cm]
Firm fixed effects&Yes&Yes\\
Year-quarter fixed effects &Yes&Yes\\	
Observations &  4,662 &  4,662 \\ 
$\text{R}^{2}$ ($\times 100$) & 14.5 & 14.5 \\ 
\bottomrule 
\end{tabular} }
\end{table} 

\newpage

\newpage

\newpage

\newpage

\newpage

\newpage

\newpage

\end{document}